\documentclass[useAMS,usenatbib]{mn2e}

\usepackage[T1]{fontenc}
 \usepackage{pslatex}
\usepackage{amsmath}
\usepackage{amsfonts}
\usepackage{color}
\usepackage{url}
\usepackage{graphicx}
\usepackage{subfigure}
\usepackage{amssymb}
\usepackage{soul}
\usepackage{booktabs}
\usepackage{array}
\usepackage{ textcomp }
\usepackage{makecell}
\usepackage{bbold}

{\newif\ifnotend
\notendtrue
\def\veclist{ABCDEFGHIJKLMNOPQRSTUVWXYZabcdefghijklmnopqrstuvwxyz.}
\def\top#1#2.{#1}
\def\tail#1#2.{#2.}
\loop\expandafter\xdef\csname v\expandafter\top\veclist\endcsname%
{{\noexpand\bf\expandafter\top\veclist}}
\edef\veclist{\expandafter\tail\veclist}
\if\veclist.\notendfalse\fi\ifnotend\repeat}
\def\Plus{\texttt{+}}
\def\Minus{\texttt{-}}

\def\pa{\partial}

\mathchardef\mhyphen="2D

\title[Periodicity makes shocks unstable]{Periodicity makes galactic shocks unstable - \rom{1}. Linear analysis}
\author[Sormani, Sobacchi, Shore, Tre{\ss} \& Klessen]{Mattia C. Sormani$^{1}$, Emanuele Sobacchi$^{2,3}$, Steven N. Shore$^{4}$, Robin G. Tre{\ss}$^1$ \newauthor and Ralf S. Klessen$^{1,5}$\\
$^1$Universit\"{a}t Heidelberg, Zentrum f\"{u}r Astronomie, Institut f\"{u}r theoretische Astrophysik, Albert-Ueberle-Str. 2, 69120 Heidelberg, Germany \\
$^2$ Physics Department, Ben-Gurion University, P.O. Box 653, Beer-Sheva 84105, Israel\\
$^3$ Department of Natural Sciences, The Open University of Israel, 1 University Road, P.O. Box 808, Raanana 4353701, Israel \\
$^4$ Dipartimento di Fisica `Enrico Fermi', University of Pisa, I-56127 Pisa, Italy\\
$^5$Universit\"at Heidelberg, Interdiszipli\"ares Zentrum f\"ur Wissenschaftliches Rechnen, Im Neuenheimer Feld 205, D-69120 Heidelberg, Germany
}
\begin{document}

\date{}

\def\p{\partial}
\def\Omegap{\Omega_{\rm p}}

\newcommand{\di}{\mathrm{d}}
\newcommand{\bfx}{\mathbf{x}}
\newcommand{\bfe}{\mathbf{e}}
\newcommand{\vlos}{\mathrm{v}_{\rm los}}
\newcommand{\Tspin}{T_{\rm s}}
\newcommand{\Tb}{T_{\rm b}}
\newcommand{\degree}{\ensuremath{^\circ}}
\newcommand{\Th}{T_{\rm h}}
\newcommand{\Tc}{T_{\rm c}}
\newcommand{\bfr}{\mathbf{r}}
\newcommand{\bfv}{\mathbf{v}}
\newcommand{\bfu}{\mathbf{u}}
\newcommand{\pc}{\,{\rm pc}}
\newcommand{\kpc}{\,{\rm kpc}}
\newcommand{\Myr}{\,{\rm Myr}}
\newcommand{\Gyr}{\,{\rm Gyr}}
\newcommand{\kms}{\,{\rm km\, s^{-1}}}
\newcommand{\kmssq}{\,{\rm km^2\, s^{-2}}}
\newcommand{\de}[2]{\frac{\partial #1}{\partial {#2}}}
\newcommand{\cs}{c_{\rm s}}
\newcommand{\rb}{r_{\rm b}}
\newcommand{\rqu}{r_{\rm q}}
\newcommand{\bfOmega}{{\Omega}}
\newcommand{\bfOmegap}{{\Omega}_{\rm p}}
\newcommand{\bfXi}{\boldsymbol{\Xi}}
\newcommand{\imi}{i}
\makeatletter
\newcommand*{\rom}[1]{\expandafter\@slowromancap\romannumeral #1@}
\makeatother

\maketitle

\begin{abstract}
We study the dynamical stability of stationary galactic spiral shocks. The steady-state equilibrium flow contains a shock of the type derived by Roberts in the tightly wound approximation. We find that boundary conditions are critical in determining whether the solutions are stable or not. Shocks are unstable if periodic boundary conditions are imposed. For intermediate strengths of the spiral potential, the instability disappears if boundary conditions are imposed such that the upstream flow is left unperturbed as in the classic analysis of D'yakov and Kontorovich. This reconciles apparently contradictory findings of previous authors regarding the stability of spiral shocks. This also shows that the instability is distinct from the Kelvin-Helmholtz instability, confirming the findings of Kim et al. We suggest that instability is a general characteristics of periodic shocks, regardless of the presence of shear, and provide a physical picture as to why this is the case. For strong spiral potentials, high post-shock shear makes the system unstable also to parasitic Kelvin-Helmholtz instability regardless of the boundary conditions. Our analysis is performed in the context of a simplified problem that, while preserving all the important characteristics of the original problem, strips it from unnecessary complications, and assumes that the gas is isothermal, non self-gravitating, non-magnetised.
\end{abstract}


\section{Introduction}

In their pioneering study, \cite{LinShu1964} already noted that the gaseous interstellar medium, given its relatively low velocity dispersion, could give rise to spiral patterns with density contrasts much stronger than the stellar counterpart. It was then demonstrated by \cite{Fujimoto1968} and \cite{Roberts1969} that the non-linear gas response to a given externally imposed rigidly rotating spiral gravitational potential can result in stationary shocks waves, provided that the amplitude of the spiral potential exceeds some critical value. These steady-state shock solutions were considered again in more detail by \cite{Shu+1973}, who studied how they depend on the underlying parameter space \citep[see also][for a historical perspective]{Toomre1977}.

A natural question arose concerning the stability of the steady state solution found by \cite{Roberts1969} and \cite{Shu+1973}. Several papers have addressed this question in the subsequent decades \citep[e.g.][]{MishurovSuchkov1975,NelsonMatsuda1977,BalbusCowie1985,Balbus1988,DwarkadasBalbus1996,LeeShu2012,LeeShu2014,KimKimKim2014,KimKimElmegreen2015}. The original calculations of \cite{Roberts1969} and \cite{Shu+1973} assumed the gas to be isothermal and non self-gravitating, but it was argued that, if any instability is present, the most likely cause would be related to the self-gravity of the gas on the basis of the high degree of compression experienced at the shocks. Hence, \cite{BalbusCowie1985} and \cite{Balbus1988} studied a self-gravitating version of the \cite{Roberts1969} problem, but found that the system was stable. Another potential source of unstable flow seemed to be related to the high shear in the post-shock region. \cite{DwarkadasBalbus1996} therefore studied the stability of the problem assuming the gas to be isothermal and non self-gravitating, exactly as in the original work of \cite{Roberts1969}, but again they found the system to be stable.

The question was revitalised by the simulations of \cite{WadaKoda2004}. These authors run simple 2D non self-gravitating simulations of isothermal gas in an externally imposed rigidly rotating spiral potential, and found that spiral shocks can be hydrodynamically unstable. They dubbed it \emph{wiggle instability} as it develops by forming ``wiggles'' along the spiral arms. They argued that it could be a manifestation of the Kelvin-Helmholtz instability due to high shear behind the shock. The instability was then seen in numerous other simulations \citep[e.g.][]{DobbsBonnell2006,Kim++2012a,KimKim2014,SBM2015a,KhoperskovBertin2015}, although \cite{HanawaKicuchi2012} suggested that it may be a numerical artefact caused by the discretisation of the fluid equations. Finally, \cite{KimKimKim2014} re-analysed the problem, assuming the gas to be isothermal and non self-gravitating exactly as in \cite{Roberts1969} and \cite{DwarkadasBalbus1996}, and this time they found the system to be unstable. They physically interpreted the instability as originating from the generation of potential vorticity at corrugated shock fronts. Other relatively recent analysis that include the effects of self-gravity and/or magnetic fields also found the solutions to be unstable \citep{LeeShu2012,KimKimElmegreen2015}.

The picture that emerges is somewhat confusing, with authors finding apparently contradictory results. Two works in particular have studied what seems to be the same problem but obtained apparently opposite results: \cite{DwarkadasBalbus1996} found the isothermal, non self-gravitating and non-magnetised spiral shocks to be stable, while \cite{KimKimKim2014} found them to be unstable. What is the cause of this discrepancy? Interestingly, the first authors assumed the upstream (with respect to the shock) flow to be unperturbed, while the second used periodic boundary conditions in their analysis. Can this difference explain the discrepancy?

A related question is the physical origin of the instability. \cite{WadaKoda2004} originally argued that the wiggle instability is essentially a Kelvin-Helmholtz instability, while \cite{KimKimKim2014} argued that the instability is physically distinct from Kelvin-Helmholtz. However, \cite{KhoperskovBertin2015} and a recent review by \cite{Shu2016} again state that it is Kelvin-Helmholtz. Is the instability of spiral shocks the same as the Kelvin-Helmholtz instability or not?

In this paper, we revisit the question of the stability in an attempt to clarify these apparently contradictory results. We reformulate the problem in a simplified context that, while preserving the important characteristics of the original problem, strips it from unnecessary complications that may obscure the analysis. Interpreting previous results in a simpler context provides physical insight into the steady state solutions and the nature of instabilities and highlights aspects of the problem that may be of a more general character.

This paper is structured as follows. In Section \ref{sec:basic} we introduce the basic equations. In Section \ref{sec:steady} we discuss the steady state background solutions. In Section \ref{sec:linear} we linearise the equations around the steady state solutions and specify the boundary conditions. In Section \ref{sec:results} we solve numerically the eigenvalue problem to find the dispersion relation and under what conditions the system is unstable. We discuss the physical interpretation of our results in Section \ref{sec:discussion} and finally summarise our conclusions in Section \ref{sec:conclusion}.

\section{Basic equations} \label{sec:basic}
\cite{Roberts1969} studied the problem of finding the gas response to a spiral stellar potential by introducing a spiral coordinate system and approximating the fluid equations in a local patch around a spiral arm under the following assumptions: i) spirals are tightly wound ii) the flow does not depend on the coordinate parallel to the spiral arm iii) the velocities induced by the spiral perturbation of the potential are small compared to the underlying circular motion of the galaxy.

Our goal is to consider the problem in the simplest possible context in order to gain physical insight into the nature of instabilities. Therefore, rather than re-deriving \cite{Roberts1969} equations, we start by studying an apparently unrelated ``toy problem'' that preserves all the important mathematical characteristics of the original problem. In Appendix \ref{appendix:correspondence} we present a derivation of \cite{Roberts1969} equations and spell out their connection with the problem considered here in the main text.

Consider a fluid in the Cartesian plane $(x,y)$ subject to the following forces:\footnote{Strictly speaking these are forces per unit mass, i.e. accelerations.} 
\begin{enumerate}
\item The pressure force,  $-{\nabla P}/{\rho}$.
\item The force from an external potential, $-\nabla \Phi$.
\item The Coriolis force, $-2 \Omega \times \bfv$. The angular velocity $\Omega$ is taken constant and directed towards the positive $z$ direction.
\item A constant force, $\mathbf{F}$. 
\end{enumerate}
The equations of motion are:
\begin{align} 
	& \pa_t \bfv + \left( \bfv \cdot \nabla \right) \bfv  = - \frac{\nabla P}{\rho} -\nabla \Phi - 2 \Omega \times \bfv + \mathbf{F} \label{eq:eom1}, \\ 
	& \pa_t \rho + \nabla \cdot \left( \rho \bfv \right) = 0 \, . \label{eq:eom2}
\end{align}
Now assume a simple externally imposed potential,
\begin{equation}
\Phi(x) = \Phi_0 \cos\left( \frac{2 \pi x}{L} \right),
\end{equation}
where $\Phi_0$ is a constant, and that the gas is isothermal,
\begin{equation}
P = \cs^2 \rho. \label{eq:isothermal}
\end{equation}

In connection with the \cite{Roberts1969} problem, these equations are meant to represent the local conditions in a patch surrounding a spiral arm, where $x$ is the coordinate perpendicular to the arm and $y$ is the coordinate parallel to the arm. The potential $\Phi$ represents the spiral perturbation to the potential (i.e. after subtraction of an underlying axisymmetric potential that is the origin of galactic circular rotation, see equations \ref{eq:split1} and \ref{eq:phiconnection}), and $L$ is the separation between two consecutive spiral arms. $\Omega$ represents the local circular speed of the galaxy (\emph{not} the pattern speed of the spiral arms, see equation \ref{eq:finaleuler} and subsequent comments). The force $\mathbf{F} = F_x \hat{e}_x + F_y\hat{e}_y$ represents the Coriolis force associated with the background circular motion, see equation \eqref{eq:Fconnection}. This is assumed to be constant which amounts to considering the circular speed constant in the local patch considered. Its components in terms of the underlying circular velocity of the galaxy are $F_x = - 2 \Omega v_{{\rm c}y}$, $F_y = 2 \Omega v_{{\rm c}x}$. Since we use the convention that $F_x,F_y>0$, the background circular flow is in the positive $x$ direction and in the negative $y$ direction in our problem. The ratio of these components is related to the pitch angle of the spiral arms by $\tan i = F_y / F_x $.

\cite{Roberts1969} \citep[see also][]{Shu+1973} showed that these equations admit steady-state solutions that are periodic in the $x$ coordinate which, if $\Phi_0$ exceeds a critical value, must contain shocks. Here, we study the linear stability of these steady-state solutions.

\subsection{Parameters counts} \label{sec:parameterscounts}

The problem posed by equations \eqref{eq:eom1} $\mhyphen$ \eqref{eq:isothermal} is completely specified by six parameters: 

\begin{equation}
\cs, \quad \Phi_0, \quad L, \quad F_x, \quad F_y, \quad \Omega.
\end{equation} 
From these, we can define 4 dimensionless parameters:
\begin{equation} \label{eq:dimensionlessparameters}
\tilde{c}_{\rm s} = \frac{\cs}{(F_y/\Omega)},\quad \tilde{\Phi}_0 = \frac{\Phi_0}{(F_y/\Omega)^2}, \;\; \tilde{L} = \frac{L}{(F_y/\Omega^2)}, \quad F_x/F_y
\end{equation}
and two ``scaling constants'' 
\begin{equation}
F_y, \quad \Omega.
\end{equation}
In what follows, without loss of generality, we assume $F_y=\Omega=1$ unless otherwise specified. We will see later that $F_x$ plays a trivial role, so the effective number of non-trivial parameters of our problem is three.

\subsection{Parameters corresponding to galactic spirals problem} \label{sec:parametersb}
Let us discuss what values of the parameters roughly correspond to the galactic spiral shocks problem. We are only interested in the orders of magnitudes rather than in precise numbers. Plausible values for the parameters are as follows. The sound speed\footnote{This is meant to be a phenomenological sound speed that takes into account in a simple way the turbulent pressure of the interstellar medium, and it is much higher than the sound speed one would obtain from the microscopic temperature of cold gas in a disk galaxy \citep[e.g.][]{Roberts1969,Cowie1980}. The ``temperature'' of the isothermal assumption is therefore related to the velocity dispersion of clouds rather than a microscopic temperature. The observed velocity dispersion of the interstellar medium seems to support this hypothesis \citep[e.g.][]{DickeyLockman1990}. In this approximation, any heating due to compression, for example at a shock, is instantaneously radiated away to restore the initial temperature.} of the interstellar medium is $\cs\simeq10\kms$ \citep[e.g.][]{Roberts1969}. The rotation speed of the Sun around the Galactic centre is $\sim~200\kms$, and the velocity perturbation due to the spiral arm potential is of the order of a few percent of the circular velocity, so we take $\Phi_0~\simeq~(10)^2 \kmssq$. The separation between two spiral arms is $L~\simeq~1 \kpc$. The angular rotation velocity of material around the Galactic centre is of order $\Omega~\simeq~20\kms\kpc^{-1}$. The constant force is about the same as the Coriolis force experienced by an object that goes at approximately the speed of the Sun,  $|\mathbf{F}|~\simeq~20\kms\kpc^{-1}\times 200\kms$. Finally, the ratio between the two components of the constant force is roughly the pitch angle of the spiral arms, which we take $F_y/F_x~\simeq~0.1$ for tightly wound spirals.

This yields the following values for the dimensionless parameters:
\begin{equation}
\cs = 0.5, \quad \Phi_0 = 0.25, \quad L = 1, \quad F_x = 10.
\end{equation}
It is interesting to note that $\cs=0.5$ is the limiting value that separates the two possible regimes (sub- or supersonic) for the $\Phi_0=0$ solution (see Section \ref{sec:trivial}). Therefore, both regimes are within plausible values of the parameters for galactic spiral shocks.

In most of the remainder of the paper, we focus on and study in detail the solutions for the following values of the parameters
\begin{equation}
L = 1, \quad \cs = 0.7 \qquad \mathrm{and} \qquad L = 1, \quad \cs = 0.3.
\end{equation}
We start considering the case $\Phi_0=0$, and then study what happens as we increase its value.

\section{Steady state} \label{sec:steady}

In this section we study steady state solutions of equations \eqref{eq:eom1} and \eqref{eq:eom2}, and in the next section we linearise the equations around these steady states. We consider steady state solutions that:
\begin{enumerate}
\item are periodic in the coordinate $x$ with the same period of $\Phi$.
\item do not depend on the coordinate $y$. 
\end{enumerate}
We obtain the following system:

\begin{align}
& v_{0y}' = -2 + \frac{1}{v_{0x}} \label{eq:s1a},\\
& v_{0x}' = \frac{ 2 v_{0y} - \Phi' + F_x}{ v_{0x} - \frac{\cs^2}{v_{0x}}}, \label{eq:s1b}
\end{align}
where the symbol $'$ denotes derivative with respect to $x$, and we used the subscript $0$ to denote the steady state solutions.
$F_x$ can be absorbed into $v_{0y}$ by means of the following transformation:
\begin{equation}
v_{0y} =  u_{0y} - \frac{F_x}{2}, \qquad \qquad v_{0x}=u_{0x}. \label{eq:t1}
\end{equation}
The equations then become
\begin{align}
& u_{0y}' = -2 + \frac{1}{u_{0x}} \label{eq:steady1}, \\
& u_{0x}' = \frac{ 2 u_{0y} - \Phi'}{ u_{0x} - \frac{\cs^2}{u_{0x}}}. \label{eq:steady2}
\end{align}
Note that while the original problem depends on four dimensionless parameters, the system of equations \eqref{eq:steady1} and \eqref{eq:steady2} depends only on three, $\cs$, $\Phi_0$ and $L$. Therefore, $u_{0x}$ and $u_{0y}$ do not depend on the fourth dimensionless parameters, $F_x$, and $v_{0x}$ and $v_{0y}$ depend on it in a trivial way. Later, we will also find that the stability of the system does not depend on $F_x$. Thus the problem has effectively three non-trivial dimensionless parameters.

\subsection{Case $\Phi_0=0$} \label{sec:trivial}

For $\Phi_0=0$, the solution to equations \eqref{eq:steady1} and \eqref{eq:steady2} is:
\begin{align}
u_{0x} =\frac{1}{2},\;\; u_{0y} = 0,
\end{align}
or, restoring the original parameters and dimensions:
\begin{align}
v_{0x} = \frac{F_y}{2 \Omega},\;\; v_{0y} = - \frac{F_x}{2 \Omega},\;\;  \rho_0 = \text{constant}. \label{eq:td}
\end{align}

There is a simple interpretation for this result. When $\Phi_0=0$, each fluid element is subject to three different forces: the Coriolis force, the constant force $\mathbf{F}$ and pressure. If the fluid element has the ``right'' velocity, the Coriolis force and the constant force $\mathbf{F}$ exactly balance, and if the fluid density is uniform, the pressure force is zero. Thus, if the fluid is moving at this equilibrium velocity and has uniform density it is in a steady state. Note also that this is essentially the geostrophic approximation for a steady imposed force \citep[e.g.][]{Pedlosky1982}.

Note also that in our dimensionless variables $u_{0x}=1/2$, regardless of the values of the other parameters, $L$ and $\cs$. Therefore, 
\begin{itemize}
\item if $\cs>1/2$, the $\Phi_0=0$ solution is subsonic,
\item if $\cs<1/2$, the $\Phi_0=0$ solution is supersonic.
\end{itemize}

As we have discussed above in Section \ref{sec:parametersb}, both these regimes are within plausible physical values for the problem of gas flowing in a spiral potential of a galaxy. Using different notation, this was already noted by \cite{Shu+1973}: in their notation the two regimes correspond to whether the Doppler-shifted phase-velocity of the stellar density wave is greater than the sound speed.

\subsection{Case $\Phi_0\neq0$}

\begin{figure*}
\includegraphics[width=\textwidth]{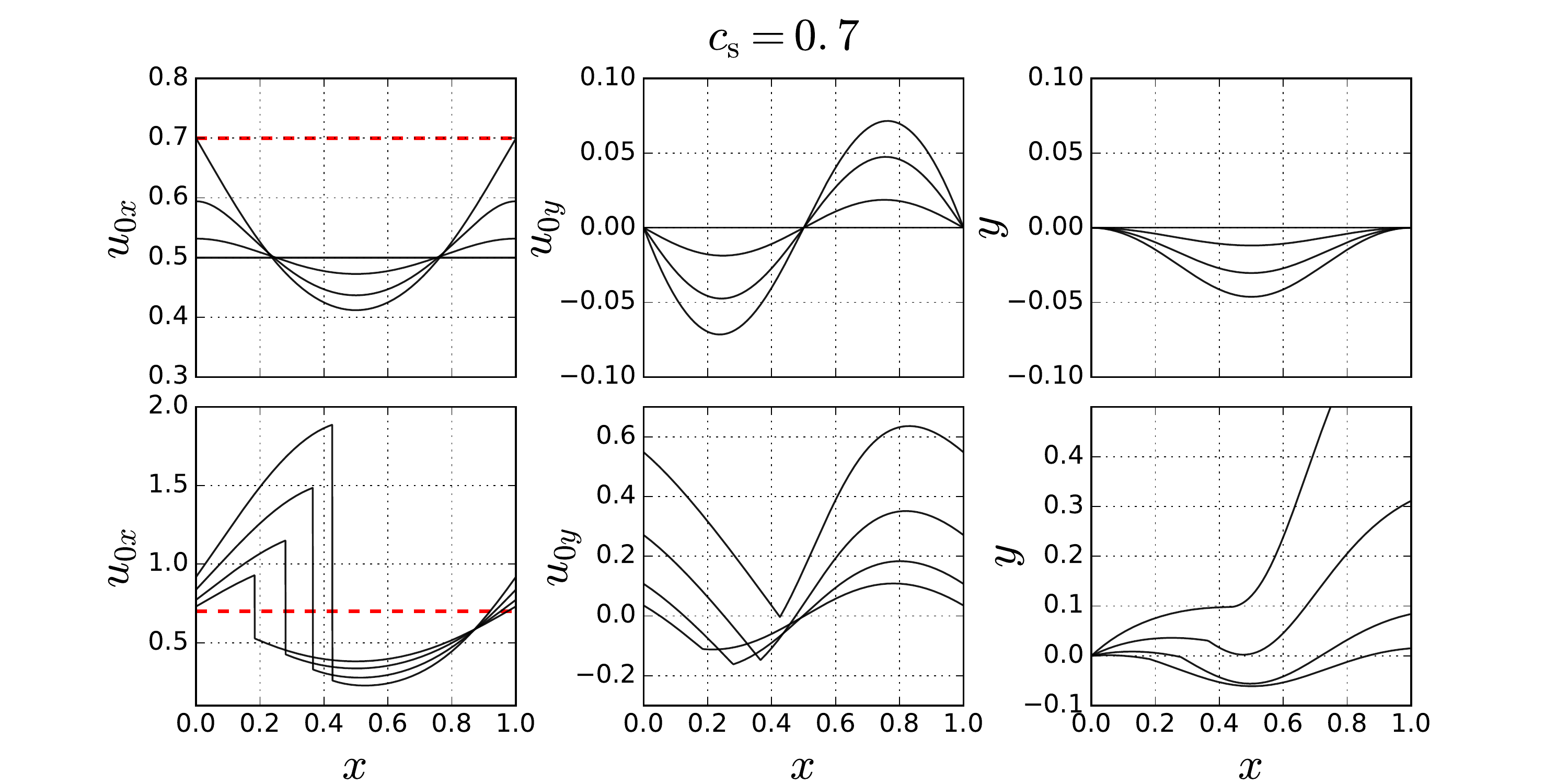}
 \caption{Steady state solutions obtained by solving Eq. \eqref{eq:steady1}$\mhyphen$\eqref{eq:steady2} for the case $L=1$, $\cs=0.7$. The top row shows cases in which $\Phi_0<\Phi_{0\rm{c}}$, when the solution does not contain a shock. Different lines correspond to $\Phi_0~=~0.00,0.02,0.05,0.07297$, leading to increasing amplitudes in $u_{0x}$ and $u_{0y}$. The bottom row shows the case $\Phi_0>\Phi_{0\rm{c}}$, when the solution does contain a shock. Different lines correspond to $\Phi_0~=~0.1,0.15,0.25,0.4$. Panels on the right show trajectories in the $(x,y)$ plane followed by fluid elements. The red dashed line is the value of the sound speed.}
\label{fig:steady07} 
\end{figure*}
\begin{figure*}
\includegraphics[width=\textwidth]{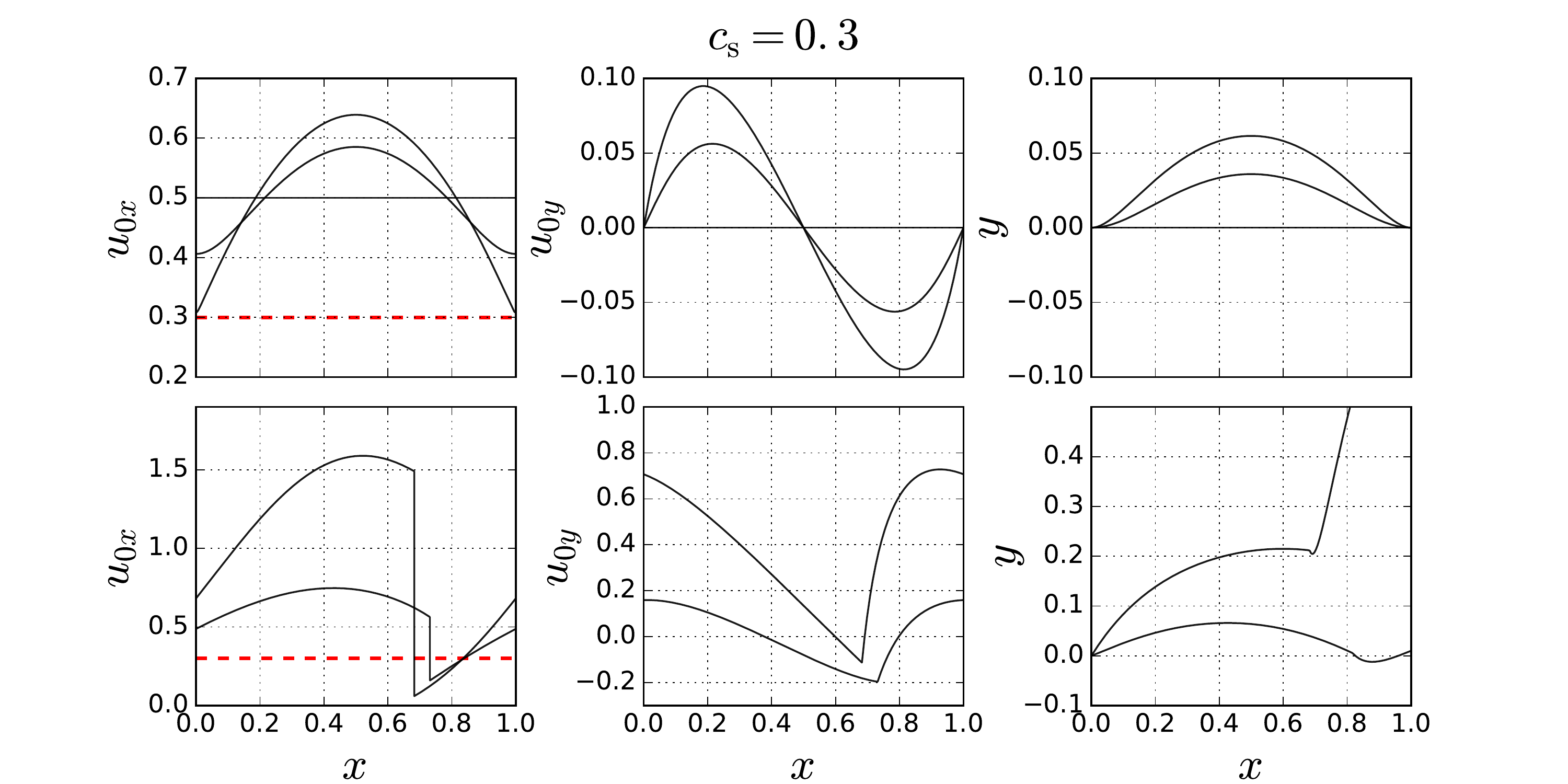}
 \caption{
Same as Fig. \ref{fig:steady07} for the case $L=1$, $\cs=0.3$. Lines in the top panels correspond to $\Phi_0~=~0.00,0.01,0.0148$, while in the bottom panels to $\Phi_0~=~0.025,0.25$.}
\label{fig:steady03} 
\end{figure*}

\subsubsection{$L=1$, $\cs=0.7$} \label{sec:caseL1cs07}

Figure \ref{fig:steady07} shows steady state solutions for the case $L=1$, $\cs=0.7$ and various values of $\Phi_0$. We have verified numerically that for each triplet $(L,\cs,\Phi_0)$, only one steady state solution exists. We also show the corresponding trajectories in the $(x,y)$ plane. 

The top row shows solutions for small values of $\Phi_0$, when a solution without a shock exists. The horizontal black full lines at $u_{0x}=0.5$ and $u_{0y}=0$ in the top-left and top-center panels respectively represent the solution for the case $\Phi=0$ discussed in Section \ref{sec:trivial}. This solution is subsonic in this case. The sound speed is represented by the red dashed line. 

When $\Phi_0$ is increased by a small amount the solutions are small amplitude librations around the $\Phi_0=0$ solution. These are the other solutions in the top row panels. When $\Phi_0\neq0$, the potential causes fluid elements to experience a varying force in the direction of motion; hence $u_{0x}$ cannot remain constant, which ``unbalances'' the Coriolis and the constant force $\mathbf{F}$ (see the discussion in Section \ref{sec:trivial}). In these steady state solutions a compromise is found, and $u_{0x}$ varies so that the Coriolis and $\mathbf{F}$ force are balanced only \emph{on average}. Indeed, we see from the top-right panel that the net displacement of a fluid element in the $y$ direction over one period is zero for small $\Phi_0$. This also means that the net energy gain of a fluid element over one period is zero.

As we increase $\Phi_0$, the amplitude of these librations grows until at some critical value, $\Phi_{\rm c}=0.07297$, the steady solution for $u_{0x}$ touches the sound speed line $\cs=0.7$. For values of $\Phi_0$ greater than this, the solution must pass from subsonic to supersonic at some point (see also the analogy with the De Laval Nozzle and Parker solar wind problem in Appendix \ref{appendix:toy}). Since we want periodic solutions, the solution needs to go back from supersonic to subsonic at some other point. This is only possible if a shock is present: therefore, for $\Phi_0>\Phi_{\rm c}$ the solution must contain a shock. 

The bottom row in Fig. \ref{fig:steady07} shows steady state solutions that contain a shock. We see from the bottom-right panel that in this case a fluid element has a net displacement in the $y$ direction over one period. Thus the fluid element drifts towards positive $y$. In the spiral arm interpretation of the problem this corresponds to a shift along the spiral arm in the opposite direction of the background flow. In a galaxy with trailing spiral arms, the direction of net transport of material to greater or smaller radii therefore depends on the relative strength of the drift (which is related to the strength of $\Phi_0$) and of the component of background circular flow parallel to the shock (which is related to $F_x$), see Eq. \eqref{eq:t1}. 

For values of $\Phi_0$ just above $\Phi_{\rm c}$, the shock appears at $x=0$, at the maximum of $\Phi$, and moves forward for increasing values of $\Phi_0$. This means that the shock is found \emph{after} the maximum of the potential, which in the spiral arm interpretation corresponds to the middle of the inter-arm region. The sonic point instead starts at $x=0$ (which is equivalent to $x=L$) and moves backwards.

It is interesting to discuss the energy balance of the system. The only force that can do a net work on the fluid over one period is $\mathbf{F}$.\footnote{The Coriolis force cannot do work since it is always directed perpendicular to the velocity of fluid elements, and the external potential returns to its initial value over one period which means there is no net gain/loss of energy due to it.} In solutions without a shock there is no net gain of energy since the net displacement in the vertical direction is zero.\footnote{Recall that we have assumed $F_x=0$ so there is no work associated with displacement in the $x$ direction.} In solutions with a shock, the fluid has a net gain of energy (proportional to $F_y\Delta y$, where $\Delta y$ is the displacement) over one period, which is then radiated away at the shock. Ultimately, this is a transfer of energy from the stellar potential that is eventually lost. The stronger $\Phi_0$, the stronger the shock, the more the energy that is radiated away at the shock, the greater must be the net $y-$displacement over one period in order to gain enough energy to compensate the higher dissipation at the shock. Stronger $y$-displacements are associated with stronger shear in the post-shock region. Thus, this explains why increasing $\Phi_0$ inevitably leads to an increase of shear in the post-shock region. 

The fact that the extra energy gained from the force $\mathbf{F}$ is radiated away is a consequence of the isothermal assumption. If we had assumed the gas to be adiabatic, so that the equations of motion satisfy conservation of energy, periodic steady states such as those considered here would not be possible. Gas would heat up steadily at each shock and the energy would be retained in the system rather than being radiated away. To prevent an ever-increasing gas temperature an external source able to subtract from the system the excess energy would be needed. Seen in a different way, this puts a limit on the cooling time of the ISM for our isothermal assumption to be valid, i.e. it must be much shorter than one period.

\subsubsection{$L=1$, $\cs=0.3$}

Figure \ref{fig:steady03} is the analog of Fig. \ref{fig:steady07} for the case $L=1$, $\cs=0.3$. The main differences from the case discussed in Section \ref{sec:caseL1cs07} are: 
\begin{enumerate} 
\item The $\Phi_0$ solution and the small amplitude solutions for small $\Phi_0$ now are supersonic rather than subsonic. Therefore, as we increase the value of $\Phi_0$ the solution touches the line $u_y=\cs$ from above rather than from below.
\item Now the shock appears at $x=0$ (equivalent to $x=L$) and moves \emph{backwards} rather than forward and is found \emph{before} the maximum of the potential ($x=0$), which corresponds to middle of the inter-arm region in the spiral arm interpretation.
\end{enumerate}
We also note that $\Phi_{\rm c}$ is significantly smaller in this case.

\section{Linear stability analysis} \label{sec:linear}

In this section we linearise the fluid equations around the steady state solutions derived in the previous section. The goal is to perform a linear stability analysis and to find the dispersion relation of the system. Since the background solution does not depend on $y$, we can write
\begin{align} 
& \rho = \rho_0\left(x\right)\left[1+ s_1\left(x\right) \exp\left(i k_y y - i \omega t \right)\right] \label{eq:rho1}, \\
& v_x = u_{0x}\left(x\right) + u_{1x}\left(x\right)\exp\left(i k_y y - i \omega t \right)  \label{eq:v1x}, \\
& v_y = -\frac{F_x}{2}+u_{0y}\left(x\right) + u_{1y}\left(x\right)\exp\left(i k_y y - i \omega t \right)  \label{eq:v1y}.\;
\end{align}
where $u_{1x}$, $u_{1y}$, $s_1$, $\omega$ are complex-valued quantities, while $u_{0x}$, $u_{0y}$, $\rho_0$, $k_y$ are real-valued quantities. Note that all quantities here are dimensionless since we have assumed $F_y=\Omega=1$ as discussed in Section \ref{sec:parameterscounts}. Plugging Eqs. \eqref{eq:rho1}-\eqref{eq:v1y} into Eqs. \eqref{eq:eom1}-\eqref{eq:eom2} and expanding to first order in the perturbed quantities (i.e., those with subscript $1$) we find:
\begin{align}
& u_{0x}\left(u_{1y}\right)'=-i k_y \cs^2 s_1+i \left( \tilde{\omega} - k_y u_{0y} \right) u_{1y}-\frac{1}{u_{0x}}u_{1x} \label{eq:main_ux}\\
& \left(u_{0x}^2-\cs^2\right)\left(u_{1x}\right)' = -i \cs^2\left( \tilde{\omega} - k_y u_{0y} \right)s_1+\left(2 u_{0x}+ik_y \cs^2\right)u_{1y} \nonumber \\
&+\left(i\left( \tilde{\omega} - k_y u_{0y} \right) u_{0x}  + \frac{ \cs^2 + u_{0x}^2}{\cs^2 - u_{0x}^2} \left(2 u_{0y} - \Phi' \right)          \right)u_{1x} \label{eq:main_uy}\\
& \left(u_{0x}^2-\cs^2\right)\left(s_1\right)'=i\left( \tilde{\omega} - k_y u_{0y} \right)u_{0x}s_1-\left(ik_y u_{0x}+2\right)u_{1y} \nonumber \\
&+\left( -i\left( \tilde{\omega} - k_y u_{0y} \right)  + \frac{2 u_{0x}}{u_{0x}^2-\cs^2} \left(2 u_{0y} - \Phi' \right)   \right)u_{1x} \label{eq:main_rho}\;,
\end{align}
where we have defined $\tilde{\omega}=\omega+k_y F_x/2$ and the symbol $'$ denotes derivative with respect to $x$. We have also used the relations \eqref{eq:steady1} and \eqref{eq:steady2} to eliminate $u_{0x}'$ and $u_{0y}'$.

The system \eqref{eq:main_ux}-\eqref{eq:main_rho} together with appropriate boundary conditions (discussed below) constitutes an eigenvalue problem. For a given $k_y$, non-trivial solutions for $u_{1x}$, $u_{1y}$ and $s_1$ (i.e., distinct from the null solution) only exist for certain discrete (but infinite in number) values of $\tilde{\omega}$. This is easy to see for example in the case $\Phi_0=0$, in which the system can be solved analytically (see Section \ref{sec:stabilitytrivial}). 

Solutions with $\operatorname{Im}(\omega)>0$ grow exponentially in time. Thus, if at least one such solution is present, the system is unstable. Solutions with $\operatorname{Im}(\omega)=0$ or $\operatorname{Im}(\omega)<0$ are respectively oscillating and damped solutions. If only solutions of these types are present, the system is stable.

Note the stability of the system does not depend on $F_x$. Indeed, $F_x$ does not appear directly in Eqs. \eqref{eq:main_ux}-\eqref{eq:main_rho}, and $u_{0x}$ and $u_{0y}$ are also independent of $F_x$. We will see below that the boundary conditions, when written in terms of $\tilde{\omega}$, are also independent of $F_x$. Hence, the eigenvalue problem for $\tilde{\omega}$ and its spectrum do not depend on $F_x$. The spectrum of $\omega$ does depend on $F_x$, but in a trivial way: changing the value of $F_x$ merely amounts to changing the real part of $\omega$, which does not affect the stability of the system. Therefore, both the steady state solutions and the spectrum of $\omega$ depend in a trivial way on $F_x$, confirming that the number of non-trivial parameters of our problem is 3 as anticipated in Sect. \ref{sec:parameterscounts}. Hereafter we assume $F_x=0$.

\subsection{Shock jump conditions}
The surface of the shock must also be perturbed. We assume that the shock front is displaced in the $x$-direction from its original location by
\begin{equation}
\xi_1=z_1\exp\left( i k_y y - i \omega t \right),
\end{equation}
where $z_1$ is a complex number and $|z_1|\ll 1$. Equations \eqref{eq:main_ux}-\eqref{eq:main_rho} are valid everywhere except at the shock surface, where we have to ensure that conservation laws\footnote{i.e., the Rankine-Hugoniot conditions.} are not violated. If we reach the shock while integrating the perturbed quantities, we have to stop using these differential equations just before the shock and perform the appropriate jump, and then use again the differential equations after the jump. 

In the reference frame of the shock, the following quantities are conserved across the shock: $\rho v_\perp$, $\left(\cs^2+v_\perp^2\right)\rho$, $v_\parallel$. Expanding to first order, the values of these quantities at the position of the perturbed shock front are:
\begin{align}
& \rho=\rho_0+\rho_1\exp\left(i k_y y - i \omega t \right)+\xi_1\frac{\text{d}\rho_0}{\text{d}x} , \\
& v_\perp=u_{0x}+u_{1x}\exp\left(i k_y y - i \omega t \right)+\frac{\text{d}u_{0x}}{\text{d}x}\xi_1+i\left( \tilde{\omega} - k_y u_{0y} \right)\xi_1 , \\
& v_\parallel=-\frac{F_x}{2}+u_{0y}+u_{1y}\exp\left(i k_y y - i \omega t \right)+\frac{\text{d}u_{0y}}{\text{d}x}\xi_1+ik_y u_{0x}\xi_1 \, .
\end{align}
At the zero-th order, the conservation laws require that $\rho_0 u_{0x}$, $\left(\cs^2+u_{0x}^2\right)\rho_0$, $u_{0y}$ are conserved across the shock; these relations are satisfied in the steady state background solutions. At first order, we find that
\begin{align}
& \rho_0 u_{0x}s_1+\rho_0 u_{1x}+i\left( \tilde{\omega} - k_y u_{0y} \right)\rho_0 z_1 , \label{eq:jump1} \\
& \frac{\cs^2+u_{0x}^2}{2u_{0x}}s_1+u_{1x} +\left(\frac{u_{0x}^2-\cs^2}{2u_{0x}^2}\frac{\text{d}u_{0x}}{\text{d}x}+i\left( \tilde{\omega} - k_y u_{0y} \right)\right)z_1, \label{eq:jump2} \\
& u_{1y}+\left(\frac{\text{d}u_{0y}}{\text{d}x}+ik_y u_{0x}\right)z_1 \label{eq:jump3} \,.
\end{align}
are conserved across the shock. Thus, equations \eqref{eq:jump1}-\eqref{eq:jump3} are the jump conditions that the perturbed quantities $u_{1x}, u_{1y}, s_1$ must satisfy at the point $x=x_{\rm sh}$, where $x_{\rm sh}$ is the position of the shock in the background steady state solution.

\subsection{Sonic point condition}

Eqs. \eqref{eq:main_ux}-\eqref{eq:main_rho} are singular at the sonic point. In other words, when the background solution satisfies $u_{0x}=\cs$, some coefficients of the differential equations diverge. To avoid divergences, the following relation must be satisfied at the sonic point:
\begin{align}
i\left(\tilde{\omega}-k_y u_{0y}\right)\cs s_1&-\left(ik_y \cs+2\right)u_{1y}\nonumber\\
&+\left[2u_{0x}'-i\left(\tilde{\omega}-k_y u_{0y}\right)\right]u_{1x}=0 \, . \label{eq:sonic}
\end{align}
This condition is obtained by plugging $u_{0x}=\cs$ in Eq. \eqref{eq:main_uy} (or equivalently in Eq. \ref{eq:main_rho}) and requiring that $u_{1x}'$ remains finite. Only if this condition is satisfied can the solution pass continuously through the sonic point. All solutions whose domain of integration contains the sonic point must satisfy this requirement.

\subsection{Boundary conditions}

Our steady state contains a shock. What boundary conditions should we impose? In their classic analysis of the corrugation instability of shock waves, \cite{Dyakov54} and \cite{Kontorovich58} leave the pre-shock flow unperturbed, on the basis of its supersonic velocity \citep[see \textsection 90 in][]{landau}. This is the correct boundary condition when we consider only a single shock. However, for \emph{sequential}, or \emph{periodic}, shocks this might not be appropriate, as material leaving one shock can later enter the next. Periodic boundary conditions seem better suited for the case of galactic spiral shocks. 

In this paper we consider both types of boundary conditions, which are described in more detail below. In the next section we explain how we implement them in our numerical code.

\subsubsection{Periodic boundary conditions}

Under periodic conditions the perturbed quantities must satisfy

\begin{align}
u_{1x}(x)& = u_{1x}(x+L)\\
u_{1y}(x) &= u_{1y}(x+L)\\
s_{1}(x) &= s_{1}(x+L)
\end{align}

\subsubsection{DK boundary conditions}

When we use D'yakov-Kontorovich (DK) boundary conditions we solve the problem only in the interval $[x_{\rm sh},x_{\rm s}]$, where $x_{\rm sh}$ and $x_{\rm s}$ indicate the position of the shock and of the sonic point respectively in the background steady state solution. At $x=x_{\rm sh}$ we assume that all pre-shock quantities are unperturbed ($u_{1x}=u_{1y}=s_1=0$ just before the shock), and all post shock quantities are such that jump conditions are satisfied accordingly. At the sonic point, we simply ask that condition \eqref{eq:sonic} is satisfied. When DK boundary conditions are imposed in this way, the flow reaches the sonic point and is able to traverse it. Since information cannot travel back after this point, it does not matter what happens after this point and we can just think of it as a sort of free-outflowing boundary.

\subsection{Numerical procedure}

We use the shooting method to solve our eigenvalue problem \citep[e.g.][]{NR2007}. Naively, one might think of shooting from an arbitrary point $x_0$. However, Eqs. \eqref{eq:main_ux}-\eqref{eq:main_rho} are singular at the sonic point. Thus if we start integrating equations \eqref{eq:main_ux}-\eqref{eq:main_rho} from a generic point $x_0$ with some random guesses as initial values, the solution will almost invariably crash at the sonic point and will not be able to traverse the entire domain of integration. Following \cite{LeeShu2012} and \cite{KimKimKim2014}, we solve this problem by using a variation of the ``shooting to a fitting point method'' \citep[e.g.][]{NR2007}: we start integrating from the sonic point, choosing initial conditions such that Eq. \eqref{eq:sonic} is already satisfied, and then integrate forward and backwards from there. We now describe in more detail our numerical procedure for our two types of boundary conditions.

\subsubsection{Numerical procedure for periodic boundary conditions}
In this case, given a value of $k_y$, we perform the following steps:
\begin{enumerate}
\item We start integrating from the sonic point $x=x_{\rm s}$ by guessing initial values for $u_{1y}$ and $\omega$. Since both are complex numbers, this amounts to guessing four real numbers. 
\item Without loss of generality we set $s_1=1+\imi$ at the sonic point (since the equations are linear, we can always perform such rescaling), and calculate $u_{1x}$ from the sonic condition \eqref{eq:sonic}.
\item We integrate backwards from $x=x_{\rm s}$ to $x=x_{\rm sh}$ and forward to $x=x_{\rm sh} + L$. 
\item We now have the values of $u_{1x}$, $u_{1y}$ and $s_1$ just before and just after the shock. We must check whether these values satisfy the jump conditions. However, we do not have a value of $z_1$ yet, as it was not necessary to start the integration from the sonic point. We use one of the jump conditions \eqref{eq:jump1}-\eqref{eq:jump3} to calculate $z_1$, and then we check whether the other two equations are satisfied. These are two complex-valued equations, so both the real and imaginary parts must be equal. This means that we have a total of 4 constraints, the same as the number of our unknowns (the 4 initial guesses). Thus the number of unknowns (initial guesses) matches the number of constraints, and we have a well defined problem. If the constraints are satisfied, we have found a good solution, if not, we have to go back and change our initial guesses (this is the essence of the shooting method).
\end{enumerate}

Thus, our numerical scheme requires essentially to find zeros of a function $\mathbb{R}^4\to\mathbb{R}^4$. To solve this problem we have used the function \emph{root} in the root finding package contained in SciPy \citep{scipy}. Different solutions are found by starting from different initial guesses. We have found that usually the solution converges to the closest available value of $\omega$.

\subsubsection{Numerical procedure for DK boundary conditions}
The procedure followed in this case is similar to the case with periodic boundary conditions. Points (i) and (ii) are the same. At point (iii), we only integrate backwards from $x=x_{\rm s}$ to $x=x_{\rm sh}$ and not forward, since in the case of DK boundary conditions we only solve the problem in the interval $[x_{\rm sh},x_{\rm s}]$. We then obtain the values of $u_{1x}$, $u_{1y}$ and $s_1$ just after the shock. Now we assume $u_{1x}=u_{1y}=s_1=0$ just before the shock, and using these values we calculate $z_1$ using one jump conditions and then check the other two complex-valued jump conditions. Thus, we again have 4 constraints and 4 unknowns, and our scheme requires finding the zeros of a function $\mathbb{R}^4\to\mathbb{R}^4$, which we solve as before.

\section{Results} \label{sec:results}

\subsection{Case $\Phi_0=0$} \label{sec:stabilitytrivial}

Let us first consider the problem with $\Phi_0=0$. When the potential vanishes, the background solution does not depend on $x$:
\begin{align}
u_{0x} = \frac{1}{2},\;\; u_{0y} = 0,\;\;  \rho_0 = \text{constant}\;.
\end{align}
Substituting $s_1\left(x\right)\propto u_{1x}\left(x\right)\propto u_{1y}\left(x\right)\propto\exp\left(ik_x x\right)$ into Eq. \eqref{eq:main_ux}-\eqref{eq:main_rho} yields a linear algebraic system. Requiring this system to have non-null solutions and restoring the original dimensions gives the following dispersion relation for sound (acoustic) waves modes:
\begin{equation}
\left(-\omega + k_x u_{0x} \right)^2 = \left( 2 \Omega \right)^2 + \cs^2 (k_x^2 + k_y^2) \;,\label{eq:disp}
\end{equation}
In addition, there are also entropy-vortex modes \citep[see for example Appendix \ref{appendix:landau} and][]{landau} which have the following dispersion relation:
\begin{equation}
k_x u_{0x} - \omega = 0. \label{eq:dispev}
\end{equation}
For both these types of modes, $\omega$ is always real (the imaginary part is zero), therefore the system is stable. 

If we consider solutions that have period $L$, then we must have:
\begin{equation}
k_x = \frac{2 \pi n}{L},  \qquad n=\{...,-1,0,1,...\}.
\end{equation}

The top panel in Fig. \ref{fig:stability1} shows the dispersion relation in the plane $(\omega,k_y)$ for the case $L=1$, $\cs=0.7$. Green horizontal lines are the entropy-vortex modes, which do not depend on $k_y$. Red and blue lines show sound waves modes obtained by taking the positive and negative square root in Eq. \eqref{eq:disp}.

\subsection{Case $\Phi_0\neq0$}

In this case we proceed as follows to find the dispersion relation in the $(\omega,k_y)$ plane. First we find modes for $k_y=0$ in the region $\operatorname{Re}(\omega)=(0,20)$. Then we follow these modes until $k_y=\pm 20$. We give tables with the spectrum for $k_y=0$ in Appendix \ref{appendix:tables}. In a few cases, we were not able to follow this modes after a certain $k_y$ due to numerical difficulties. However we have also manually explored the parameter space up to $|k_y|=100$ and $|\operatorname{Re}({\omega})|=100$ to check whether any conclusion that we have drawn was changed by exploring a larger region, and we found that our conclusions are unaffected. 

\begin{figure*}
\includegraphics[width=\textwidth]{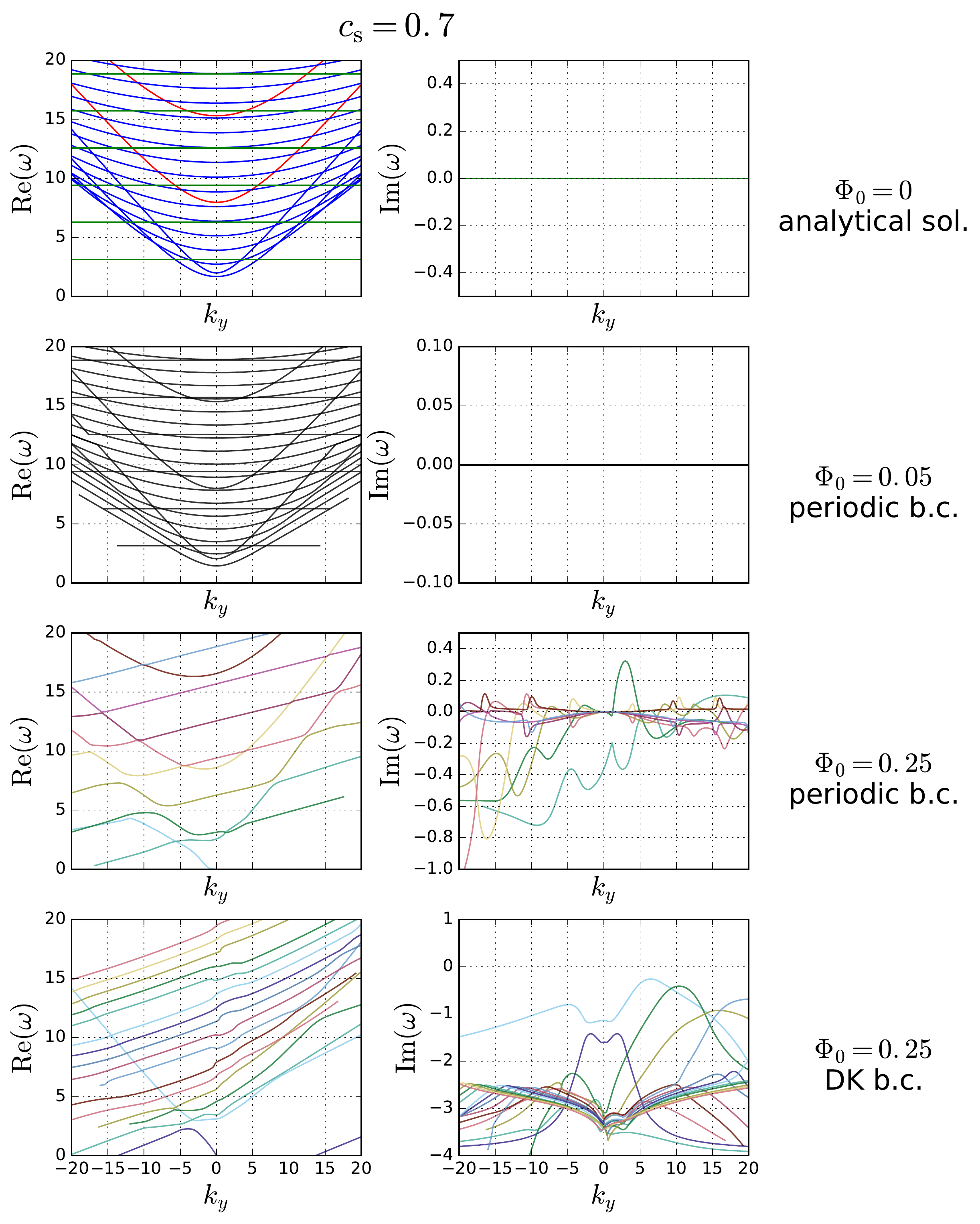}
 \caption{Dispersion relations for four different cases with $L=1$, $\cs=0.7$. The first row shows the dispersion relation for $\Phi_0=0$ calculated analytically from eqs \eqref{eq:disp} and \eqref{eq:dispev}. The other rows show the dispersion relation calculated numerically for the cases $\Phi_0=0.05$ and periodic boundary conditions (which does not contain a shock), $\Phi_0=0.25$ with periodic boundary conditions and $\Phi_0=0.25$ with DK boundary conditions. The corresponding steady state background solutions are shown in Fig
  \ref{fig:steady07}. Only the case $\Phi_0=0.25$ with periodic boundary conditions has unstable modes. The colour coding in the left and right panels correspond for the same modes.}
\label{fig:stability1} 
\end{figure*}

Figure \ref{fig:stability1} shows the dispersion relation for four different cases with $L=1$, $\cs=0.7$. 
\begin{enumerate}
\item The first row is the case $\Phi_0=0$, already discussed in the previous section. 
\item The second row shows the case $\Phi_0=0.05$ with periodic boundary conditions, for which the background solution does not contain a shock.\footnote{The numerical procedure followed in this case is similar to the case with shocks but more straightforward, since the background solution does not contain a shock nor a sonic point the equations are not singular anymore.} As one would expect, in this case the dispersion relation is very similar to that for $\Phi_0=0$ and the system is stable. Some entropy-vortex modes seem to stop beyond a certain $k_y$ in the figure, but this is likely a numerical artefact of our code as these modes were sometimes difficult to follow numerically without jumping onto some other mode.
\item The third row shows the more interesting case $\Phi_0=0.25$ with periodic boundary conditions. Here the dispersion relation is more complicated. There are unstable modes, with $\operatorname{Im}(\omega)>0$, and damped modes, with $\operatorname{Re}(\omega)<0$. The system is unstable.
\item The fourth row shows the case $\Phi_0=0.25$ with DK boundary conditions. The only difference between this and the previous case are the boundary conditions. Here, only damped modes exist and the system is stable. Changing the boundary conditions has made the instability disappear.
\end{enumerate}

\begin{figure*}
\includegraphics[width=\textwidth]{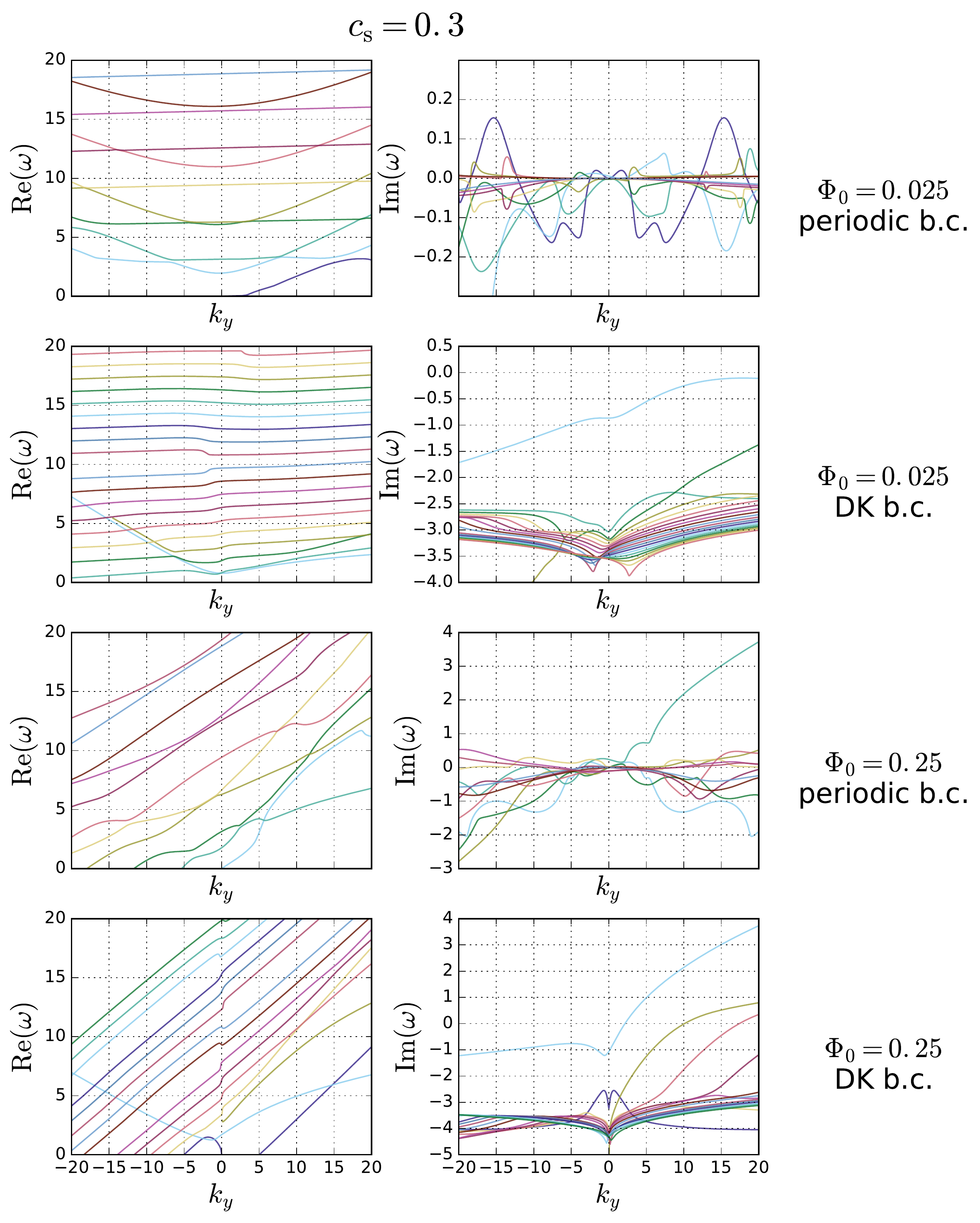}
 \caption{Dispersion relations for four different cases with $L=1$, $\cs=0.3$. The first and second rows show the cases $\Phi_0=0.025$ with periodic and DK boundary conditions respectively. In this case, changing the boundary conditions makes the unstable modes disappear. The third and fourth rows show the cases  $\Phi_0=0.25$ with periodic and DK boundary conditions respectively. This time, changing the boundary conditions does not make the instability disappear. The corresponding steady state background solutions are shown in Fig. \ref{fig:steady03}. The colour coding in the left and right panels correspond for the same modes.}
\label{fig:stability2} 
\end{figure*}

Figure \ref{fig:stability2} shows the dispersion relation for four different cases with $L=1$, $\cs=0.3$. 
\begin{enumerate}
\item The first row is the case $\Phi_0=0.025$ with periodic boundary conditions. This case is unstable, similarly to the case $\Phi_0=0.25$ with periodic boundary conditions in Fig. \ref{fig:stability1}.
\item The second row is the same case with DK boundary condition. There are only damped modes. Again, changing the boundary condition makes the instability disappear.
\item The third row shows the more interesting case $\Phi_0=0.25$ with periodic boundary conditions. Now the system is extremely unstable. $\operatorname{Im}(\omega)$ reaches values much higher than in the previous cases (which means the instability develops much faster) and peaks at higher values of $k_y$. Also note that the most unstable mode has two ``bumps''.
\item The fourth row shows the case $\Phi_0=0.25$ with DK boundary conditions. This time the instability does not disappear changing the boundary conditions. The system is again extremely unstable. The most unstable mode is similar to the previous case but this time has only one ``bump''.
\end{enumerate}

\section{Discussion} \label{sec:discussion}

\subsection{Physical interpretation} \label{sec:interpretation}

Table \ref{table:summary} summarises our results. We have found that for moderate values of $\Phi_0$ (i.e. when $\Phi_0$ is not too far from $\Phi_{\rm c}$) changing the boundary conditions makes the instability disappear. Only with periodic boundary conditions the system is unstable, while it is stable for DK boundary conditions. For stronger values of $\Phi_0$ instead, the system is unstable regardless of the boundary conditions. What is the physical reason behind this behaviour?

In the D'yakov-Kontorovich classic analysis, isothermal shocks are always found to be stable \citep[see Appendix \ref{appendix:landau} and \textsection 90 in][]{landau}. In their analysis the upstream flow is assumed to be unperturbed because of its supersonic velocity (any perturbation is advected with the flow and eventually disappears if it is not maintained by an external forcing), and only the shock surface and the downstream flows are perturbed. Stable modes exist with some characteristic frequencies, and the shock surface can oscillate with these frequencies (see our equation \ref{eq:resultDK}). However, if one modifies their problem to send incident waves from upstream towards the shock, these can resonate with the natural frequencies of the shock, which are the stable oscillating modes found by DK. Thus, if one sends incident waves from upstream with the right frequencies, it is possible to make the shock resonate and blow up (see Appendix \ref{appendix:landau}).

Crucially, and unlike in the DK case where they disappear forever, in the periodic case waves coming out from one shock can enter into the next. This suggests that these waves can excite resonant modes, eventually leading to instability. Hence, while single shocks are generally stable as shown by DK and subsequent authors, periodic shocks are generally unstable because they ``resonate with themselves''. This scenario is realised in a disk galaxy, where material coming out from a spiral arm can enter the next spiral arm. We note that this behaviour might have applications to other contexts in which periodic shocks are present, such as tidally-induced shocks in accretion disks in close binary systems.

This interpretation is complementary to that of \cite{KimKimKim2014}. These authors argued that potential vorticity is generated at each passage at a deformed shock front, while it is conserved between two shocks. The continuous amplification of the potential vorticity by subsequent shocks leads to instability. 

Our results also explain why \cite{DwarkadasBalbus1996} found the system to be stable, while \cite{KimKimKim2014} using what is seemingly the same setup found it to be unstable. The former used DK boundary conditions, leading to a stable system, while the latter used periodic boundary conditions, leading to an unstable system. \cite{KimKimKim2014} have noted this discrepancy but have attributed it to poor numerical resolution and an insufficiently long integration time in the work of \cite{DwarkadasBalbus1996}. We suggest instead that the discrepancy is caused by the different boundary conditions. More generally, a careful examination of the literature shows that all works finding a stable system \citep{BalbusCowie1985,Balbus1988} use boundary conditions akin to DK, in the sense they do not allow material that leaves one shock to enter into the next shock, while works that find the system to be unstable \citep{LeeShu2012,KimKimKim2014,KimKimElmegreen2015} use periodic boundary conditions. Although some of these studies included self-gravity and magnetic field, it may be that the key effect leading to instability is whether shocks are considered to be periodic or not. We argue that shocks are essentially periodic in a real disk galaxy, so the instability must appear there (see Section \ref{sec:reallyperiodic}). The morphology of the resulting ``feathering'' may depend on the details of the physics included, but the presence of such ``feathering'' may ultimately be attributable to material passing through a succession of shocks. 

The fact that the instability disappears by switching from periodic to DK boundary conditions also confirms that the instability is distinct from a Kelvin-Helmholtz instability (KHI) as argued by \cite{KimKimKim2014}. Indeed, if the instability was caused by a KHI due to shear in the post-shock region, it would depend on the local conditions after the shock only and would not be affected by a change in the boundary conditions. 

When $\Phi_0$ is increased, we have noted that the instability no longer disappears by changing the boundary conditions. In this case, a parasitic KHI due to high shear in the post-shock region appears along the periodic shock instability discussed above. This is not surprising given how the shear in the post-shock region increases as we increase $\Phi_0$: a glance at the bottom-middle panel of Fig. \ref{fig:steady03} shows that when $\Phi_0$ is increased from $0.025$ to $0.25$ the background solution for $u_{0y}$ is much steeper in the post-shock region, triggering a true KHI. We have checked that eventually the same happens if we increase $\Phi_0$ in the case $\cs=0.7$. The timescale for the parasitic KHI to develop is usually much shorter than the periodic shock instability: the latter cannot be too fast because it requires fluid elements to complete at least a one period to become effective. Also note that the most unstable mode in the third row of Fig. \ref{fig:stability2} has a double bump, while it has a single bump in the fourth row. This suggests that the first bump is due to the periodic shock instability, while the second bump is the parasitic KHI. The timescales associated with the two bumps seem to confirm this.

To further test that our overall interpretation is correct, we have done two things. First, we have been investigated the problem also using hydrodynamical simulations. These have confirmed our results and will be the subject of a companion paper. Preliminary results suggest that the linear stability analysis can predict accurately the timescales and wavelengths of the instability. Second, we have considered an even simpler toy problem than the one discussed so far. In this toy problem, the steady state solution contains periodic shocks similar to the previous case but $u_{0y}=0$ for all values of $\Phi_0$. If our picture is correct, we should find that this system is always unstable with periodic boundary conditions while it is always stable with DK boundary conditions, regardless of the strength of $\Phi_0$. It should not be possible to trigger the parasitic KHI in this toy problem as post-shock shear is never present. This is indeed what we found. This simpler toy problem is described in Appendix \ref{appendix:toy}.

\subsection{Dependence on the parameters}

We have found that the stability of the system does not depend on $F_x$. Since $F_y/F_x=\tan i$ (see Section \ref{sec:basic}), this means that, fixed the values of all other parameters, the stability is independent of the pitch angle. This seems at odd with the results of \cite{WadaKoda2004}, who find that the stability is sensitive to the pitch angle, see their figure 7. This apparent contradiction is explained if we consider that varying only the pitch angle in the models of \cite{WadaKoda2004} actually corresponds to varying many parameters simultaneously in our models. In particular, varying only the pitch angle in \cite{WadaKoda2004} amounts to varying both $F_x$ and $F_y$ in our models while keeping $|\mathbf{F}|$ constant. Since our dimensionless variables are scaled according to $F_y$ (equation \ref{eq:dimensionlessparameters}) this amounts to varying all our dimensionless parameters, which should be all varied simultaneously for a fair comparison with the simulations of \cite{WadaKoda2004}. Moreover, varying the pitch angle in \cite{WadaKoda2004} also corresponds to significantly changing the interarm distance $L$ in our models. Finally, our equations are strictly valid in the tightly wound approximation, which is not valid in the right panel in figure 7 of \cite{WadaKoda2004}, and this may lead to further differences.

Our point is that, at least in the tightly wound approximation, larger pitch angles do not necessarily correspond to more unstable systems, but one must also be careful to specify which other parameters are kept constant in the analysis.

Our results also indicate that the instability is stronger at lower sounds speed. This is expected because lower velocity dispersions lead to stronger density contrasts and greater Mach numbers in response to a given spiral potential and is in agreement with the findings of \cite{WadaKoda2004} and \cite{KimKimKim2014}. Note that changing the value of the sound speed while keeping constant all the other parameters in our or in the cited references yields a fair comparison, so the interpretation is easier than in the case of the pitch angle.

According to Table \ref{table:summary}, greater Mach numbers correlate with systems that are more unstable to KHI. Note however that this is a case of ``correlation does not mean causation'', and there is only an indirect causal connection between the Mach number and the occurence of KHI. The latter is caused by shear. As discussed in Section 3, in our models the amount of shear is connected to the shock strength, because stronger shocks dissipate more energy which must be compensated with greater displacements in the vertical directions, which amounts to more shear. Hence, the shear and the Mach number are correlated because they have a common origin in these models, i.e. they both depend on the strength of the spiral potential. But this is only because the Coriolis term mixes the $x$ and $y$ direction. When shear is absent, as in the toy problem considered in appendix D, arbitrarily high Mach number do not lead to KHI. In this sense there is no direct connection between a higher Mach number and the occurrence of a KHI.

It would be interesting to understand how the wavelength of the most unstable mode and the threshold that marks the occurrence of the KHI depend on the parameters $(L,\Phi_0,\cs)$. However, a systematic exploration of parameter space is cumbersome to do with the linear analysis, while simple hydro simulations are more suited to this task. Therefore, we plan to carry out a wider exploration of the parameter space in the previously mentioned companion paper. 

\subsection{Are galactic shocks really periodic?} \label{sec:reallyperiodic}

In our models, as in real galaxies, streamlines are not strictly closed (see the right-bottom panels of Figures \ref{fig:steady07} and \ref{fig:steady03} and the discussion of ``drifts'' in Section \ref{sec:steady}). However, our models are translationally invariant in the $y$ direction (i.e. the direction parallel to a spiral arm), while in a real galaxy (and also in the simulations of \citealt{WadaKoda2004}) the flow smoothly changes as we move along a spiral arm. Thus one may ask whether our results apply to a real galaxy and/or to global hydrodynamical simulations. 

We argue on the basis of the physical interpretation discussed in Section \ref{sec:interpretation} that although not exactly periodic, the underlying physical picture based on amplification of small disturbances is still valid and should lead to instability. Small disturbances coming out from one particular shock will subsequently hit a slightly different shock at a different radius, but they will nevertheless be amplified because the transmission coefficients are usually very high (see also Appendix \ref{appendix:landau}). Figure 2 and 5 in work of \cite{KimKim2014} seem to confirm this. In these figures, one sees that the instability developing from the centre outwards. Our results suggest that this is because it is in the centre that the time separation between two shock passages is shortest, hence the disturbances are amplified earliest there. The same figures also seem to indicate that ``wiggle'' and ``ripples'' form where perturbations are coming into the shock. Indeed, large amplification of inhomogeneities was noted by \cite{DobbsBonnell2006} in their simulations as gas goes through a spiral shock. The typical timescales derived from the linear analysis show that a small number of passages are enough to get into the non-linear regime. With hindsight, this is not surprising given the analysis in Appendix \ref{appendix:landau}, which shows that in principle a wave sent with exactly the right frequency, regardless of how it originated, can resonate and result in an infinite amplification factor. Therefore it is likely to be unimportant where the original disturbances are coming from, or whether from an identical shock or a slightly different one. 

A more subtle question is how to separate, in a real galaxy or in a simulation like those of \cite{WadaKoda2004}, the contributions from the periodic shock instability and the parasitic KHI. While in our idealised problem it is possible to turn off the former by changing the boundary conditions, this is not possible in a real galaxy and in global simulations. This is why the idealised studies are useful, because they allow study of physical mechanisms from a privileged point of view, in this case by isolating two effects that are otherwise difficult to separate. However, \emph{the KHI timescale is usually much shorter than the periodic instability timescale}. Therefore, if in a simulation like those of \cite{WadaKoda2004} an instability develops \emph{before} the gas had time to cover the distance between two spiral arms, then it must be a KHI, while if it develops over times longer than this, it is likely to be a periodic shock instability. Note however that even in the case a parasitic KHI is present, once the disturbances it creates reach the next shock they will be greatly amplified \citep[see also][]{DobbsBonnell2006}. It is ultimately possible that in real galaxies both processes, the periodic shock instability and the KHI, are simultaneously operating and that what has been called ``wiggle instability'' is a combination of both precesses.

\begin{table}
\caption{Summary of the stability results. $M=u_{0x}/\cs$ is the Mach number calculated just before the shock.} \label{table:summary}
\begin{center}
\begin{tabular}{ c c c c c } 
boundary conditions & $\Phi_0$ & $\cs$ & $M$ & Stability \\
 \hline
Periodic & 0.25 & 0.7 & 2.1 & Unstable \\
DK & 0.25 & 0.7 & 2.1 & Stable  \\
Periodic & 0.025 & 0.3 & 1.9 & Unstable \\
DK & 0.025 & 0.3 & 1.9 & Stable  \\
Periodic & 0.25 & 0.3 & 5.0 & Unstable \\
DK & 0.25 & 0.3 & 5.0 & Unstable  \\
 \hline
\end{tabular}
\end{center}
\end{table}

\section{Conclusion} \label{sec:conclusion}

We used a linear stability analysis to study the stability of stationary galactic spiral shocks. The steady-state equilibrium flow contains a shock of the type first derived by Roberts in the tightly wound approximation. We have found that the occurrence of an instability depends crucially on the boundary conditions imposed. Our analysis is performed in the context of a simplified problem in order to make the physical interpretation of the results as clear as possible. We have also assumed that gas is isothermal, non self-gravitating, non-magnetised. We have found that:
\begin{enumerate}
\item Galactic shocks are always unstable when periodic boundary conditions are imposed.
\item For moderate strengths of the spiral potential, the instability disappears if boundary conditions are switched to those used in the classic works of \cite{Dyakov54} and \cite{Kontorovich58} in which the upstream flow is left unperturbed. 
\item The key physical motivation that leads to instability in the periodic case is that small amplitude sound waves and entropy-vortex waves leaving one shock can enter into the next shock, be amplified and resonate with it, leading to instability. This type of periodic shock instability is what has been previously called ``wiggle instability''. Based on this physical interpretation, we have argued that instability is a general characteristic of periodic shocks, even outside the galactic shocks context.
\item The periodic shock instability is not a parasitic a Kelvin-Helmholtz instability due to shear in the post-shock region, otherwise it would not disappear by changing the boundary conditions. This explains apparently contradictory findings in the literature and suggests that periodic shocks might be the key to understand the feathering of spiral arms. Self-gravity and/or magnetic fields are certainly important in determining the morphology of feathers but they may not be the primary driver.
\item For higher strengths of the spiral potential, the shear in the post-shock region must increase as an inevitable consequence of shocks getting stronger: stronger shocks dissipate more energy which must be compensated by a larger drift in the vertical direction, i.e. more shear (see Section \ref{sec:caseL1cs07}). Parasitic Kelvin-Helmholtz instabilities can develop in this case on top of the periodic shock instability.
\end{enumerate}
Our analysis is strictly valid only in the tightly wound approximation, but we have argued on the basis of our physical interpretation that mechanism that leads to the instability should be applicable whenever disturbances can be amplified by a sequence of shocks. The results obtained by a linear stability analysis in this paper have been confirmed by hydrodynamical simulations which will be the subject of a companion paper which is currently in preparation.

\section*{Acknowledgements}
The authors thank Steve Balbus, Giuseppe Bertin, James Binney, Kees Dullemond, Simon Glover, John Magorrian, Kiwan Park, Matthew Ridley and the anonymous referee for useful comments and discussions and Woong-Tae Kim for kindly helping with the numerical procedure. MCS acknowledges support from the Deutsche Forschungsgemeinschaft in the Collaborative Research Center (SFB 881) ``The Milky Way System'' (subprojects B1, B2, and B8) and in the Priority Program SPP 1573 ``Physics of the Interstellar Medium'' (grant numbers KL 1358/18.1, KL 1358/19.2). We furthermore thank the European Research Council for funding in  the ERC Advanced Grant STARLIGHT (project number 339177). ES acknowledges support from the Israeli Science Foundation under Grant No. 719/14. SNS thanks the Astronomical Institute, Charles Univ., for a visiting professorship.

\def\aap{A\&A}\def\aj{AJ}\def\apj{ApJ}\def\mnras{MNRAS}\def\araa{ARA\&A}\def\aapr{Astronomy \&
  Astrophysics Review}\def\apjs{ApJS}
\bibliographystyle{mn2e}
\bibliography{bibliography}

\appendix

\section{Derivation of basic equations} \label{appendix:correspondence}
\subsection{Equations of motion in a rotating frame}
The Euler and continuity equations in a frame rotating with pattern speed $\bfOmegap$ are:

\begin{align} 
	& \pa_t \rho + \nabla \cdot \left( \rho \bfv \right) = 0 , \label{eq:eom_1} \\
	& \pa_t \bfv + \left( \bfv \cdot \nabla \right) \bfv  = - \frac{\nabla P}{\rho} -\nabla \Phi - 2 \bfOmegap \times \bfv - \bfOmegap \times \left( \bfOmegap \times \bfr \right) , \label{eq:eom_2}
\end{align}
where $\bfv$ is the velocity in the rotating frame, $- 2 \bfOmegap \times \bfv$ is the Coriolis force, $ - \bfOmegap \times \left( \bfOmegap \times \bfr \right)$ is the centrifugal force.

\subsection{Spiral coordinates}
\label{sec:spiral_coord}
\begin{figure}
\includegraphics[width=0.5\textwidth]{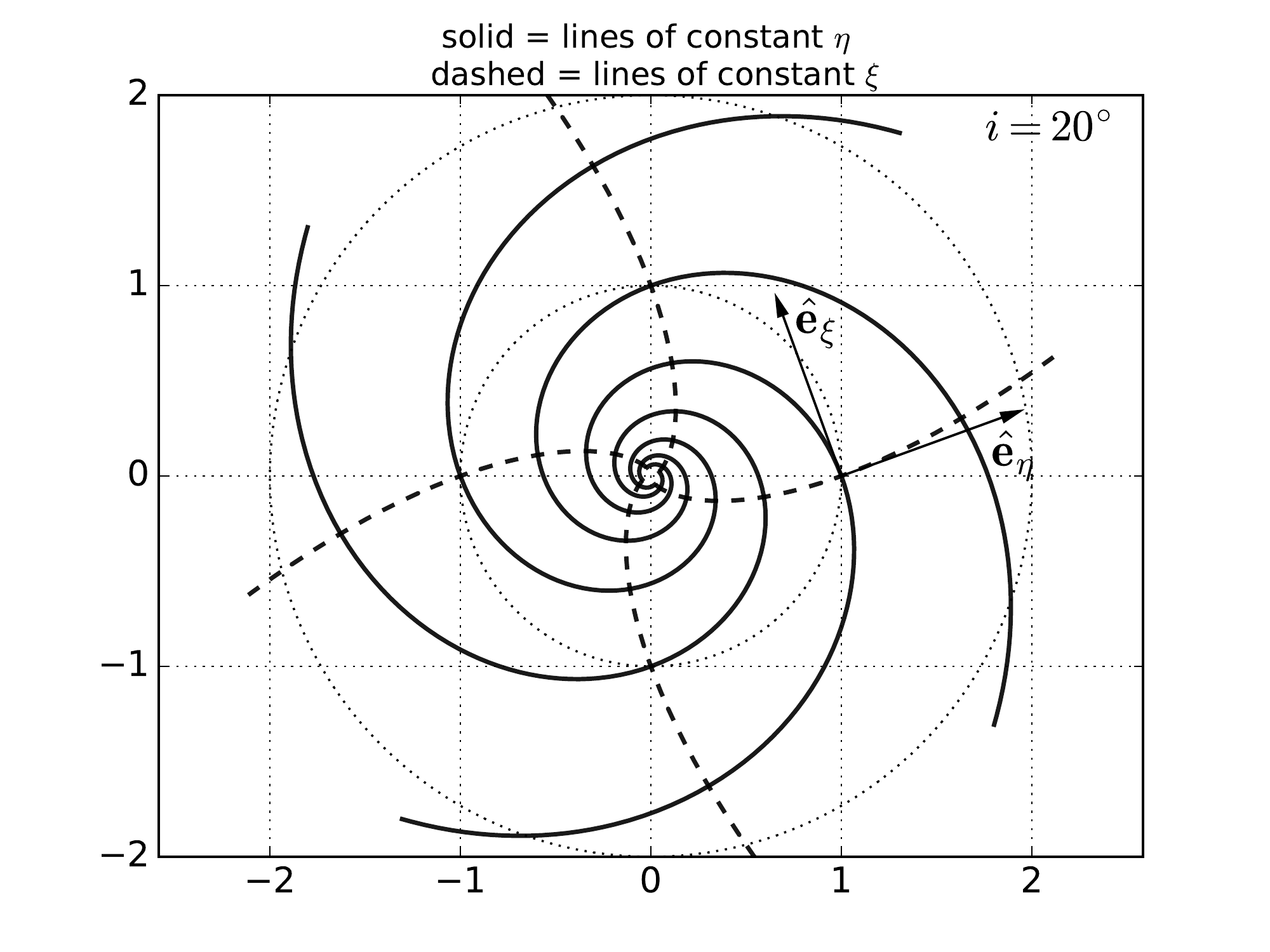}
 \caption{Definition of the spiral coordinate system.}
\label{fig:spiralcoord} 
\end{figure}

Following \cite{Roberts1969}, we define the following spiral coordinates:
\begin{align}
 \eta & = \log\left(R/R_0\right) \cos(i)+\theta\sin(i), \\
 \xi   & = -\log\left(R/R_0\right) \sin(i)+\theta\cos(i).
\end{align}
The inverse relations are:
\begin{align}
 \log\left(R/R_0\right) = \eta \cos(i) - \xi \sin(i), \\
 \theta = \eta \sin(i)+ \xi \cos(i),
\end{align}
where $R$, $\theta$ are usual polar coordinates and $R_0$ and $i$ are constants. Fig. \ref{fig:spiralcoord} shows lines of constant $\eta$ and $\xi$. The unit vectors in the directions $\eta$ and $\xi$ are:
\begin{align}
\hat{e}_{\eta} & = \cos(i) \hat{e}_{R} + \sin(i) \hat{e}_{\theta}, \\ 
\hat{e}_{\xi}   & = - \sin(i) \hat{e}_{R} + \cos(i) \hat{e}_{\theta}.
\end{align}
Straightforward calculations show that the gradient in spiral coordinates is:
\begin{equation}
{\nabla}  =  \frac{1}{R}\left( \hat{e}_{\eta} \de{}{\eta} + \hat{e}_{\xi} \de{}{\xi} \right)
\end{equation}
and the derivatives of the unit vectors are:
\begin{align}
\de{\hat{e}_{\eta}}{\eta} & = \sin(i) \hat{e}_{\xi}, \qquad  & \de{\hat{e}_{\eta}}{\xi} &= \cos(i) \hat{e}_{\xi}, \\
\de{\hat{e}_{\xi}}{\eta}   & = -\sin(i) \hat{e}_{\eta}, \qquad & \de{\hat{e}_{\xi}}{\xi} &= - \cos(i) \hat{e}_{\eta} .
\end{align}

\subsection{Equations of motion in spiral coordinates}
Using the relations of the previous subsection it is straightforward to rewrite the fluid equations \eqref{eq:eom_1} and \eqref{eq:eom_2} in spiral coordinates. The continuity equation becomes:
\begin{equation} 
	\pa_t \rho + \frac{1}{R}\left[ \pa_{\eta} \left(\rho v_{\eta}\right) + \pa_{\xi} \left(\rho v_{\xi}\right) + \rho \left(v_{\eta} \cos(i) - v_{\xi} \sin(i) \right)\right] = 0,
\label{eq:continuity}
\end{equation}
and the Euler equation:
\begin{align}
	\pa_t v_{\eta} + &\frac{1}{R} \left[ v_{\eta} \left(\pa_{\eta} v_{\eta}\right) + v_{\xi} \left(\pa_{\xi} v_{\eta}\right) - v_{\xi} v_{\eta} \sin(i) - v_{\xi}^2 \cos(i) \right] \nonumber \\
		& = - \frac{1}{R} \frac{ \pa_{\eta} P}{\rho} - \frac{1}{R} \pa_{\eta} \Phi + 2 \Omegap v_{\xi} +\cos(i) \Omegap^2, \\
	\pa_t v_{\xi}   + &\frac{1}{R} \left[ v_{\eta} \left(\pa_{\eta} v_{\xi}\right) + v_{\xi} \left(\pa_{\xi} v_{\xi}\right) + v_{\eta} v_{\xi} \cos(i) + v_{\eta}^2 \sin(i) \right] \nonumber \\ 
		& = - \frac{1}{R} \frac{ \pa_{\xi} P}{\rho} - \frac{1}{R} \pa_{\xi} \Phi - 2\Omegap v_{\eta} -\sin(i) \Omegap^2 .
\end{align}

\subsection{Split into circular and spiral components}
Consider an axisymmetric steady-state solution of the fluid equations such that: 
\begin{itemize}
\item A completely axisymmetric background potential $\Phi_0$ is present.
\item The gas is in purely circular motion with velocity 
\begin{equation}
\bfv_{\rm c} = \left( \bfOmega(R) - \bfOmegap \right) \times \bfr.
\end{equation}
\item The density $\rho_{\rm c}$ is uniform.
\end{itemize}
Such a solution satisfies the following equations:
\begin{align} 
	\pa_t \rho_{\rm c} + \nabla \cdot \left( \rho_{\rm c} \bfv_{\rm c} \right) & = 0, \label{eq:eom_axi1}\\
	\pa_t \bfv_{\rm c} + \left( \bfv_{\rm c} \cdot \nabla \right) \bfv_{\rm c} & = - \frac{\nabla P_{\rm c}}{\rho_{\rm c}} -\nabla \Phi_0 - 2 \bfOmegap \times \bfv_{\rm c} - \bfOmegap \times \left( \bfOmegap \times \bfr \right), \label{eq:eom_axi2}\\
	\pa_t \rho_{\rm c} & = \pa_t \bfv_{\rm c} = \nabla P_{\rm c} = 0. \label{eq:eom_axi3}
\end{align}
Now add a spiral component $\Phi_{\rm s}$ to the external potential. We write all fluid quantities as the sum of the axisymmetric solution plus a ``spiral'' departure from the axisymmetric solution. Hence we write
\begin{equation}
\begin{split} \label{eq:split1}
& \bfv = \bfv_{\rm c} + \bfv_{\rm s}\\
& \rho = \rho_{\rm c} + \rho_{\rm s}\\
& P = P_{\rm c} + P_{\rm s} \\
& \Phi = \Phi_0 + \Phi_{\rm s}
\end{split}
\end{equation}
Substituting \eqref{eq:split1} into \eqref{eq:eom_1}$\mhyphen$\eqref{eq:eom_2} and using \eqref{eq:eom_axi1}-\eqref{eq:eom_axi3} to eliminate some terms we find: 
\begin{align}
	& \pa_t \rho_{\rm s} + \nabla \cdot \left( \rho \bfv \right)  = 0 ,\\
	& \pa_t \bfv_{\rm s} + \left( \bfv \cdot \nabla \right) \bfv_{\rm s} + \left( \bfv_{\rm s} \cdot \nabla \right) \bfv_{\rm c}   = - \frac{\nabla P_{\rm s}}{\rho} -\nabla \Phi_{\rm s} - 2 \bfOmegap \times \bfv_{\rm s}. \label{eq:cont+euler}
\end{align}
Note that so far we have not performed any approximation.

\subsection{Approximation}

Following \cite{Roberts1969} (see also \citealt{Balbus1988}) we now approximate the equations of motion in a local patch centred at a radius $R_0$ under the following assumptions:
\begin{itemize}
\item The pitch angle is small,
\begin{equation} \tan i \ll 1. \end{equation}
\item The circular speed $R \Omega$ is of the same order of $v_\xi$ and is much greater than $v_\eta$, $v_{{\rm s}\xi}$, $v_{{\rm s}\eta}$. The latter are all comparable in size. Thus
\begin{equation} v_\eta \sim v_{{\rm s}\eta} \sim v_{{\rm s}\xi} \ll R \Omega \sim v_\xi. \end{equation}
\item The radial spacing between the spiral arms $L$ is much smaller than $R_0$.
\begin{equation} L \ll R. \end{equation} 
\item Quantities vary much faster in the direction $\hat{e}_\eta$ (with a length-scale $L$), while they vary more slowly (with a length-scale $R$) in the direction $\hat{e}_\xi$. Thus
\begin{equation} \pa_\eta \sim (R/L), \quad \pa_\xi \sim 1. \end{equation} 
\end{itemize}

\subsubsection{The continuity equation}
Consider Eq. \eqref{eq:continuity}. Using the approximations listed in the previous subsection, we see that:
\begin{align}  
\pa_{\eta} \left(\rho v_{\eta}\right) & \sim (R/L) \rho v_{\eta}, \\ 
\pa_{\xi} \left(\rho v_{\xi}\right) &\sim \rho v_{\xi}, \\
\rho v_{\eta} \cos(i) &\sim \rho v_{\eta}, \\
 \rho v_{\xi} \sin(i) &\sim  \rho v_{\xi} \sin(i).
 \end{align}
 The last two quantities are negligible compared to the first two. Hence we can approximate the continuity equation as:
\begin{equation}
\pa_t \rho + \frac{1}{R}\left[ \pa_{\eta} \left(\rho v_{\eta}\right) + \pa_{\xi} \left(\rho v_{\xi}\right) \right] = 0 \;.
\end{equation}
If we now focus on the neighbour of a point at distance $R=R_0$ from the centre of the galaxy and define $x$ and $y$ coordinates such that $\text{d}x=R_0\text{d}\eta$ and $\text{d}y=R_0\text{d}\xi$, at first order we find
\begin{equation}
 \pa_t \rho + \left[ \pa_{x} \left(\rho v_{x}\right) + \pa_{y} \left(\rho v_{y}\right) \right] = 0. \label{eq:finalcontinuity}
\end{equation}

\subsubsection{The Euler equation}
Consider the following identity:
\begin{equation} \label{eq:identity}
\left( \bfv_{\rm s} \cdot \nabla \right) \bfv_{\rm c} = \left( \bfOmega - \bfOmegap \right) \times \bfv_{\rm s} + \frac{\bfv_{\rm s} \cdot \bfr }{R} \left[ \left( \frac{\di \bfOmega}{\di R} \right) \times \bfr \right].
\end{equation}
We can expand and approximate the second term on the right-hand side of Eq. \eqref{eq:identity} and obtain
\begin{align}
\frac{\bfv_{\rm s} \cdot \bfr }{R} \left[ \left( \frac{\di \bfOmega}{\di R} \right) \times \bfr \right] & = R \frac{\di \Omega}{\di R} \left( v_{{\rm s} \xi} \cos(i) - v_{{\rm s} \eta} \sin(i) \right)  \left( \cos(i) \hat{e}_{\xi} + \sin(i) \hat{e}_{\eta} \right) \\
& \simeq R \frac{\di \Omega}{\di R} v_{{\rm s} \xi}  \hat{e}_{\xi},
\end{align}
where we have used that $\tan i \ll 1$ and that in spiral coordinates we have
\begin{equation}
\bfr = R \left( \cos(i) \hat{e}_{\eta} - \sin(i) \hat{e}_{\xi} \right).
\end{equation}
We can therefore approximate Eq.~\eqref{eq:identity} as:
\begin{equation}
\label{eq:vdotnablavc2}
\left( \bfv_{\rm s} \cdot \nabla \right) \bfv_{\rm c}= \left( \bfOmega - \bfOmegap \right) \times \bfv_{\rm s}+R \frac{\di \Omega}{\di R} v_{{\rm s} \xi}  \hat{e}_{\xi} \;.
\end{equation}
Using the relations of Section \ref{sec:spiral_coord} and that $\tan i \ll 1$ we find
\begin{align}
\left( \bfv \cdot \nabla \right) \bfv_{\rm s} &= \frac{1}{R}\left[\left(v_\eta\pa_\eta v_{\rm s\eta}+v_\xi\pa_\xi v_{\rm s\eta}\right)\hat{e}_{\eta}+\left(v_\eta\pa_\eta v_{\rm s\xi}+v_\xi\pa_\xi v_{\rm s\xi}\right)\hat{e}_{\xi}\right] \nonumber \\
& +\frac{1}{R}\left(-v_\eta v_{\rm s\xi}\sin\left(i\right)-v_\xi v_{\rm s\xi}\cos\left(i\right)\right)\hat{e}_\eta \nonumber \\
&+ \frac{1}{R}\left(v_\eta v_{\rm s\eta}\sin\left(i\right)+v_\xi v_{\rm s\eta}\cos\left(i\right)\right)\hat{e}_\xi  \\
& \simeq \frac{1}{R}\left[\left(v_\eta\pa_\eta v_{\rm s\eta}+v_\xi\pa_\xi v_{\rm s\eta}\right)\hat{e}_{\eta}+\left(v_\eta\pa_\eta v_{\rm s\xi}+v_\xi\pa_\xi v_{\rm s\xi}\right)\hat{e}_{\xi}\right] \nonumber \\
& +\frac{1}{R}\left[-v_\xi v_{\rm s\xi}\hat{e}_\eta+ v_\xi v_{\rm s\eta}\hat{e}_\xi\right] \;.
\end{align}
Note that some terms in the above equations arise from the derivatives of the unit vectors $\hat{e}_\eta$ and $\hat{e}_\xi$. Since $v_\xi\simeq v_{\rm c\xi}=\left(\Omega-\Omega_{\rm p}\right)R\cos\left(i\right)\simeq\left(\Omega-\Omega_{\rm p}\right)R$, this equation can be rewritten as
\begin{align} 
\left( \bfv \cdot \nabla \right) \bfv_{\rm s} & = \frac{1}{R}\left[\left(v_\eta\pa_\eta v_{\rm s\eta}+v_\xi\pa_\xi v_{\rm s\eta}\right)\hat{e}_{\eta}+\left(v_\eta\pa_\eta v_{\rm s\xi}+v_\xi\pa_\xi v_{\rm s\xi}\right)\hat{e}_{\xi}\right]\nonumber \\
& + \left( \bfOmega - \bfOmegap \right) \times \bfv_{\rm s}\;. \label{eq:vdotnablavs}
\end{align}
Substituting Equations \eqref{eq:vdotnablavc2} and \eqref{eq:vdotnablavs} into Equation \eqref{eq:cont+euler} and defining $x$ and $y$ coordinates such that $\text{d}x=R_0\text{d}\eta$ and $\text{d}y=R_0\text{d}\xi$, we finally find
\begin{align}
\pa_t \bfv_{\rm s} &+  \left(v_{\rm x}\pa_{\rm x} v_{\rm sx}+v_{\rm y}\pa_{\rm y} v_{\rm sx}\right)\hat{e}_{\rm x}+\left(v_{\rm x}\pa_{\rm x} v_{\rm sy} +v_{\rm y}\pa_{\rm y} v_{\rm sy}\right)\hat{e}_{\rm y} \nonumber \\
& =- \frac{\nabla P_{\rm s}}{\rho} -\nabla \Phi_{\rm s} - 2 \bfOmega \times \bfv_{\rm s}-R_0 \frac{\di \Omega}{\di R} v_{\rm sy}  \hat{e}_{\rm y} . \label{eq:finaleuler}
\end{align}
which agrees with the result of \cite{Roberts1969}, \cite{Balbus1988} and \cite{KimKimKim2014}. In this equation, ${\di \Omega}/{\di R}$ is calculated at the point $R_0$. Note that
\begin{itemize} 
\item in equation \eqref{eq:finaleuler} the derivatives do not act on the unit vectors.
\item the Coriolis term that appears in this equation is \emph{not} calculated using the pattern speed $\Omega_{\rm p}$, but using the value of $\Omega$ at $R=R_0$, i.e. of the angular rotation speed of the galaxy relative to an inertial frame. However, the total velocities in the same equation are calculated in the frame that rotates with $\Omega_p$.
\item in solving equations \eqref{eq:finalcontinuity} and \eqref{eq:finaleuler}, the circular velocity $\bfv_{\rm c}$ must be specified. Variation of $v_{{\rm c}x}$ and $v_{{\rm c}y}$ as a function of $x$ and $y$ give rise to terms whose magnitude is comparable to other terms in equations \eqref{eq:finalcontinuity} and \eqref{eq:finaleuler}, therefore the circular velocity cannot in general be considered constant \emph{independently} from the form of the function $\Omega(R)$. 
\end{itemize}

\subsection{Connection with the problem considered in the main text}

Consider equations \eqref{eq:finalcontinuity} and \eqref{eq:finaleuler}. In these equations, the total velocity is given by
\begin{align} 
v_x &= v_{{\rm c}y} + v_{{\rm s}x}, \\ 
v_y &= v_{{\rm c}y} + v_{{\rm s}y}.
\end{align}
Following \cite{BalbusCowie1985}, \cite{DwarkadasBalbus1996} and others let us assume that the circular velocity can be considered constant and equal to
\begin{align} 
v_{{\rm c}x} &= \left(\Omega(R_0)-\Omega_{\rm p}\right)R_0 \sin(i), \\
v_{{\rm c}y} &= \left(\Omega(R_0)-\Omega_{\rm p}\right)R_0 \cos(i),
\end{align}
then the various terms in equation \eqref{eq:finaleuler} can be rewritten as follows: 
\begin{align}
& \pa_t \bfv_{\rm s}  =  \pa_t \bfv, \\ 
& v_{\rm x}\pa_{\rm x} v_{\rm sx}+v_{\rm y}\pa_{\rm y} v_{\rm sx}   =  v_{\rm x}\pa_{\rm x} v_{\rm x}+v_{\rm y}\pa_{\rm y} v_{\rm x}, \\
& v_{\rm x}\pa_{\rm x} v_{\rm sy} +v_{\rm y}\pa_{\rm y} v_{\rm sy} = v_{\rm x}\pa_{\rm x} v_{\rm y} +v_{\rm y}\pa_{\rm y} v_{\rm y}, \\
& \frac{\nabla P_{\rm s}}{\rho}  = \frac{\nabla P}{\rho}, \\ 
& 2 \bfOmega \times \bfv_{\rm s} = 2 \bfOmega \times ( \bfv - \bfv_{\rm c} )
\end{align}
Hence, we can rewrite \eqref{eq:finaleuler} as:
\begin{align}
\pa_t \bfv &+  \left(v_{\rm x}\pa_{\rm x} v_{\rm x}+v_{\rm y}\pa_{\rm y} v_{\rm x}\right)\hat{e}_{\rm x}+\left(v_{\rm x}\pa_{\rm x} v_{\rm y} +v_{\rm y}\pa_{\rm y} v_{\rm y}\right)\hat{e}_{\rm y} \\
& =- \frac{\nabla P}{\rho} -\nabla \Phi_{\rm s} - 2 \bfOmega \times \bfv +  2 \bfOmega \times \bfv_{\rm c} - R \frac{\di \Omega}{\di R} v_{\rm sy} \hat{e}_{\rm y} \label{eq:mainfinal}
\end{align}
Equations \eqref{eq:eom1} and \eqref{eq:eom2} can be obtained from equations \eqref{eq:finalcontinuity} and \eqref{eq:mainfinal} provided that i) the term $\di \Omega / \di R$ is neglected. This simplifies the problem conceptually while not affecting the important mathematical characteristics of the problem nor the conclusions in the main text; ii) the following change of notation is performed:
\begin{equation}  \label{eq:Fconnection}
\mathbf{F} = 2 \bfOmega \times \bfv_{\rm c},
\end{equation}
and
\begin{equation} \label{eq:phiconnection}
\Phi = \Phi_{\rm s}.
\end{equation}

\section{Steady states for $\Phi_0\ll1$. Resonances.}

In the main text we have studied numerically exact solutions of equations \eqref{eq:steady1} and \eqref{eq:steady2}. However, when $\Phi_0\ll1$ is very small, it is possible to find approximate steady state solutions analytically by expanding the equations to first order in small quantities and thus recover the small amplitude solutions without shocks found in the main text. This also shows that depending on the values of the parameters it is possible for resonances to occur, for which the gas response to the imposed potential is particularly strong \citep{Shu+1973}.

Beginning with the solution for the case $\Phi_0=0$ discussed in Section \ref{sec:trivial}:
\begin{align}
u_{0x} =\frac{1}{2},\;\; u_{0y} = 0,
\end{align}
we look for solutions to equations \eqref{eq:steady1} and \eqref{eq:steady2} which are close to the $\Phi_0=0$ solution, 
\begin{align}
u_{0x} =  \frac{1}{2} + \Delta u_{0x},\;\; u_{0y} = \Delta u_{0y}. \label{eq:r1}
\end{align}
Substituting \eqref{eq:r1} into \eqref{eq:steady1} and \eqref{eq:steady2} and expanding to first order in the quantities with $\Delta$ and in $\Phi$, we obtain:
\begin{equation}
\Delta u_{0x}'' = - \frac{16}{\left(1 - 4 \cs^2 \right)} \Delta u_{0x} - \frac{2}{\left(1 - 4 \cs^2 \right)}\Phi'' \label{eq:forced}
\end{equation}
The solution of eq. \eqref{eq:forced} with period $L$ is:\footnote{Note that, if $\cs<0.5$, this is the equation of a driven harmonic oscillator, where the driving force is given by the term with $\Phi''$.}
\begin{equation} \label{eq:l2}
\Delta u_{0x} =  \left(\frac{2 \pi}{L}\right)^2 \frac{2 \Phi_0}{16 - \left(1 - 4 \cs^2\right)(2 \pi / L)^2} \cos\left( \frac{2 \pi x}{L} \right)
\end{equation}
This is the approximate steady-state analytical solution for the case of small $\Phi_0$. Note that the denominator diverges when
\begin{equation}
16 - (1 - 4 \cs^2) (2 \pi / L)^2  = 0. \label{eq:c1}
\end{equation} 
This result has a simple interpretation. Consider small amplitude sound waves propagating through the $\Phi_0~=~0$ solution, when the medium has uniform background density and uniform background velocity $\bfv_0$. The dispersion relation for these waves is:
\begin{equation}
\left(- \omega + \mathbf{k} \cdot \bfv_0 \right)^2 = \left( 2 \Omega \right)^2 + \cs^2 \mathbf{k}^2. \label{eq:dr}
\end{equation}
Therefore, the trajectory of a fluid element in these sound waves has the following form:
\begin{equation}
\bfx(t) = \bfv_0 t + \bfx_1 \exp\left( i \mathbf{k} \cdot \bfx - i \sqrt{ \left( 2 \Omega \right)^2 + \cs^2 \mathbf{k}^2} \; t\right).  \label{eq:traj}
\end{equation}
In other words, a fluid element propagates oscillating around a straight line with frequency
\begin{equation}
 \tilde{\omega} = \sqrt{ \left( 2 \Omega \right)^2 + \cs^2 \mathbf{k}^2}.
\end{equation} 
When $\Phi_0\ll1$, sound waves are similar to those for the case $\Phi_0=0$, and a fluid element following the trajectory \eqref{eq:traj} encounters the maxima of the gravitational potential at time intervals separated by $L/v_{0x} = {2 \Omega L}/{F_y} $. Hence, the fluid element feels a \emph{periodic} external forcing due to the gravitational potential with frequency:
\begin{equation}
\omega_\Phi = \frac{\pi F_y}{ \Omega L}.
\end{equation} 
According to the dispersion relation \eqref{eq:dr}, the frequency of sound waves with wavelength equal to the distance between maxima of the potential is ($k_x=2 \pi / L$, $k_y = 0$):
\begin{equation}
 \tilde{\omega} = \sqrt{ \left( 2 \Omega \right)^2 + \cs^2 \left( \frac{2 \pi}{L}\right) ^2}.
\end{equation} 
The condition for resonances is
\begin{equation}
 \tilde{\omega} = \omega_\Phi,
 \end{equation}
which is precisely the same as the condition \eqref{eq:c1}, after the original parameters with dimensions are restored. In other words: resonances are present when sound-wave oscillations of a fluid element have the same frequency as the forcing caused by the external potential on the same fluid element. 

Retaining more terms in the expansions in the quantities $\Delta u_{0x}$ and $\Delta u_{0y}$ leads to higher order resonances \citep{Shu+1973}.

\section{The D'yakov-Kontorovich analysis} \label{appendix:landau}
In this appendix we briefly review some of the classic results contained in a series of papers by \cite{Dyakov54,Dyakov58a,Dyakov58b} and \cite{Kontorovich58,Kontorovich58b}. In the earlier works \citep{Dyakov54,Kontorovich58}, these authors considered the stability of a single planar shock wave to corrugations of its surface in a fluid with an arbitrary equation of state. In this analysis, only the downstream flow is perturbed, while the upstream fluid is assumed to be unperturbed because of its supersonic velocity, which would advect any perturbation to infinity if not maintained by an external forcing. They found that shocks are generally stable, except for exotic equations of state \citep[see also the discussion in \textsection 90 of][]{landau}. 

In later works (\citealt{Dyakov58a,Dyakov58b,Kontorovich58b}, see also \citealt{McKenzieWestphal1968}), these authors studied the transmission and reflection coefficients for small amplitude perturbations (waves) coming from upstream or downstream with an arbitrary angle of incidence. There are two possible kinds of small perturbations that can propagate in a uniform fluid moving with constant velocity: sound waves, which move with the speed of sound relative to the fluid, and entropy-vortex waves that are simply advected with the fluid \citep[see for example \textsection 82 in][]{landau}. When a wave of only one type is incident on the shock,\footnote{Note that any small amplitude perturbations can be uniquely decomposed as a linear superposition of sound and entropy-vortex waves, so it suffices to study the transmission and reflection of each type separately.} the waves that diverge from the shock are generally composed of both types.\footnote{Note that since entropy-vortex waves are advected with the flow, they can only hit the shock from upstream, not from downstream} When a wave is incident from upstream, only transmitted waves can be present, and no reflected wave, since the upstream flow is supersonic. On the other hand, when a wave is incident from downstream, only a reflected wave is present, for the same reason. Both transmission and reflection can result in great amplification of the disturbances. 

Here, we adapt the more general results obtained by the authors mentioned above to our simpler isothermal case. We begin with a recap of the properties of sound and entropy-vortex waves.
\subsection{Sound waves and entropy-vortex waves}
There are two possible kinds of small perturbations in a uniform fluid moving with constant velocity: sound waves and entropy-vortex waves . For our isothermal case, these can be characterised as follows. Let us denote with $s_1\equiv \rho_1/\rho_0$ the density perturbation and with $\bfv_1$ the velocity perturbation. 

For a sound wave
\begin{align}
\bfv_1& = \bfv_1 ^{(\rm s)} \exp\left( i \mathbf{k} \cdot \bfx - i \omega t \right), \label{eq:sound1} \\
s_1    &= s_1 ^{(\rm s)} \exp\left( i \mathbf{k} \cdot \bfx - i \omega t \right), \label{eq:sound2}
\end{align}

where the dispersion relation and velocity perturbation are

\begin{align}
 \cs^2 \mathbf{k}^2  & =  \left( \omega - \mathbf{k} \cdot \bfv_0 \right)^2,  \label{eq:sound3} \\
\bfv_1 ^{(\rm s)} & = \cs^2 \frac{\mathbf{k}}{ ( \omega - \mathbf{k}\cdot \bfv_0 )} s_1 ^{(\rm s)}. \label{eq:sound4} 
\end{align}
Therefore, a sound wave is completely specified by three quantities: $\omega$, $k_y$ and $ s_1 ^{(\rm s)}$. The latter specifies the amplitude of the wave.

For an entropy vortex wave
\begin{align}
\bfv_1& = \bfv_1 ^{(\rm e)} \exp\left( i \mathbf{k} \cdot \bfx - i \omega t \right), \label{eq:ex1} \\
s_1    &= s_1 ^{(\rm e)} \exp\left( i \mathbf{k} \cdot \bfx - i \omega t \right), \label{eq:ex2}
\end{align}

where the dispersion relation and velocity perturbation are

\begin{align}
\mathbf{k} \cdot \bfv_0 & = \omega, \label{eq:ex3} \\
\mathbf{k} \cdot  \bfv_1 ^{(\rm e)}  & = 0, \label{eq:ex4} \\
 s_1 ^{(\rm e)} & = 0. \label{eq:ex5}
\end{align}
Thus, an entropy-vortex wave is also completely specified by three quantities: $\omega$, $k_y$ and $ v_{1x} ^{(\rm e)}$. The latter specifies the amplitude of the wave.

\subsection{Stability of isothermal shocks and their interaction with small perturbations incident from upstream}
Consider a planar shock wave at rest at $x=0$ in a fluid that obeys the following equations of motion:
\begin{align} 
	& \pa_t \bfv + \left( \bfv \cdot \nabla \right) \bfv  = - \cs^2 \frac{\nabla \rho}{\rho} \label{eq:eomdk1}, \\ 
	& \pa_t \rho + \nabla \cdot \left( \rho \bfv \right) = 0 \, . \label{eq:eomdk2}
\end{align}
The unperturbed flow is assumed to move from left to right (i.e., in the positive $x$ direction). The upstream density and speed are assumed to be:
\begin{align}
\rho_0^{(\Minus)}& = \alpha \rho_0, \\
v_0^{(\Minus)}& =  v_0/\alpha,
\end{align}
and for the downstream flow:
\begin{align}
\rho_0^{(\Plus)} &= \rho_0, \\
v_0^{(\Plus)} &= v_0,
\end{align}
where we have defined 
\begin{equation}
\alpha \equiv \left( \frac{v_0}{\cs}\right)^2,
\end{equation}
and we have $\alpha<1$ since the downstream flow must be subsonic.
We take both the upstream (-) and downstream (+) flow to be a superposition of sound waves and entropy vortex waves:
\begin{align}
s_1^{(\pm)} & = \left[ s_1 ^{(\rm e \pm)} \exp\left( i k_x^{(\rm e \pm) } x\right)  + s_1 ^{(\rm s \pm)}  \exp\left( i k_x^{(\rm s \pm)  } x \right)  \right] \exp\left(ik_y y - i \omega t \right) , \\ 
\bfv_1^{(\pm)} & = \left[ \bfv_1 ^{(\rm e \pm)} \exp\left( i k_x^{(\rm e \pm) } x\right)  + \bfv_1 ^{(\rm s \pm)}  \exp\left( i k_x^{(\rm s \pm)  } x \right)  \right] \exp\left(ik_y y - i \omega t \right)\;.
\end{align}
This encompasses both the case in which the upstream flow is unperturbed (which corresponds to $\bfv_1^{(\Minus)} = s_1^{(\Minus)}=0$) and the case in which small perturbations are incident from upstream.\footnote{The upstream perturbations are waves that are assumed to be sent from $x=-\infty$ by an external source.} Note that $k_y$ must be the same upstream and downstream, while $k_x$ is general different.\footnote{The change in the ratio $k_x/k_y$ is directly related to reflection and refraction laws similar to Snell's law in geometrical optics, see for example \cite{McKenzieWestphal1968}.} Using the conditions \eqref{eq:ex3}-\eqref{eq:ex5} and \eqref{eq:sound3}-\eqref{eq:sound4} we can write:
\begin{align}
s_1 ^{(\rm e \pm)}  & = 0, \\
k_x^{(\rm e \pm)} & = \omega / v_0^{(\pm)} ,\\
k_x^{(\rm s \pm)} & \equiv k_x^{(\pm)} , \\
 \cs^2 \left( (k_x^{(\pm)})^2 + k_y^2 \right) & =  \left( \omega -k_x^{(\pm)} v_0^{(\pm)} \right)^2 , \label{eq:dispdk1}\\
\bfv_1 ^{(\rm s \pm)} & =  \cs^2 \frac{\mathbf{k}^{(\pm)}}{ ( \omega -  k_x^{(\pm)} v_0^{(\pm)})} s_1 ^{(\rm s \pm)} , \\
\bfv_1 ^{(\rm e \pm)} & = v_{1x}^{(\rm e \pm)} \left( \hat{\mathbf{x}} - \frac{\omega}{k_y v_0^{(\pm)}} \hat{\mathbf{y}} \right) .
\end{align}
Thus at $x=0$ we have:
\begin{align}
s_1^{(\pm)} & = s_1 ^{(\rm s \pm)}, \label{eq:s1pm} \\
v_{1x}^{(\pm)} & = v_{1x}^{(\rm e \pm)} + \cs^2 \frac{{k_x}^{(\pm)}}{ ( \omega - k_x^{(\pm)} v_0^{(\pm)})} s_1 ^{(\rm s \pm)}, \label{eq:v1xpm}  \\
v_{1y}^{(\pm)} & = - v_{1x}^{(\rm e \pm)} \frac{\omega}{k_y v_0^{(\pm)}}   + \cs^2 \frac{{k_y}}{ ( \omega - k_x^{(\pm)} v_0^{(\pm)})} s_1 ^{(\rm s \pm)}. \label{eq:v1ypm}
\end{align}
As in the main text, the surface of the shock is assumed to be displaced in the $x$ direction by an amount:
\begin{equation}
\xi_1 = z_1 \exp(i k_y y - i \omega t),
\end{equation}
where $z_1$ is a complex number. We can now use the jump conditions \eqref{eq:jump1}-\eqref{eq:jump2} to relate the downstream quantities to the upstream quantities. For the case considered in this appendix these can be written as:
\begin{align}
& \rho_0^{(\Minus)}  v_{1x}^{(\Minus)} s_1^{(\Minus)} + \rho_0^{(\Minus)}  v_{1x}^{(\Minus)} + i \omega \rho_0^{(\Minus)}  z_1 =  \rho_0^{(\Plus)}  v_{1x}^{(\Plus)} s_1^{(\Plus)} + \rho_0^{(\Plus)}  v_{1x}^{(\Plus)} + i \omega \rho_0^{(\Plus)} z_1,  \label{eq:jumpdk1} \\
& \frac{\cs^2+\left[v_0^{(\Minus)} \right]^2}{2 v_0^{(\Minus)} }s_1^{(\Minus)} +v_{1x}^{(\Minus)} = \frac{\cs^2+\left[v_0^{(\Plus)} \right]^2}{2 v_0^{(\Plus)} }s_1^{(\Plus)} +v_{1x}^{(\Plus)}  ,\label{eq:jumpdk2} \\
& v_{1y}^{(\Minus)}+ i k_y v_{0}^{(\Minus)} z_1 =  v_{1y}^{(\Plus)}+ i k_y v_{0}^{(\Plus)} z_1. \label{eq:jumpdk3} \,
\end{align}
Plugging equations \eqref{eq:s1pm}-\eqref{eq:v1ypm} into equations \eqref{eq:jumpdk1}-\eqref{eq:jumpdk3} we obtain the following system:
\begin{equation}
\mathbb{A} \mathbf{\mathrm{X}} = \mathbb{b}, \label{eq:dyakov1}
\end{equation}
where 
\begin{equation}
\mathbf{\mathrm{X}} =  \begin{pmatrix}  s_1 ^{(\rm s \Plus)} \\ v_{1x}^{(\rm e \Plus)}  \\ z_1 \end{pmatrix},
\end{equation}
\begin{equation}
\mathbb{A} =  \begin{pmatrix} v_0 + \frac{k_x^{(\Plus)} \cs^2}{\omega - k_x^{(\Plus)} v_0} & 1 & i \left[1 - \frac{v_0^2}{\cs^2}\right] \omega \\ \frac{\cs^2+v_0^2}{2 v_0} + \frac{k_x^{(\Plus)} \cs^2}{\omega - k_x^{(\Plus)} v_0} & 1 & 0 \\ \frac{k_y \cs^2}{\omega - k_x^{(\Plus)} v_0} & -\frac{\omega}{k_y v_0} & i k_y v_0 \left[ 1 - \frac{\cs^2}{v_0^2}\right] \end{pmatrix} ,
\end{equation}
and
\begin{equation} \label{eq:dyakov}
 \mathbb{b}  =  \begin{pmatrix} v_0 + \frac{k_x^{(\Minus)} v_0^2}{\omega - k_x^{(\Minus)} \cs^2/v_0} & \frac{v_0^2}{\cs^2} \\  \frac{\cs^2+v_0^2}{2 v_0} + \frac{k_x^{(\Minus)} \cs^2}{\omega - k_x^{(\Minus)} \cs^2/v_0} & 1 \\ \frac{k_y \cs^2}{\omega - k_x^{(\Minus)} \cs^2/v_0} & -\frac{\omega v_0}{k_y \cs^2}\end{pmatrix} 
  \begin{pmatrix}  s_1 ^{(\rm s \Minus)} \\ v_{1x}^{(\rm e \Minus)}  \end{pmatrix} .
\end{equation}
Equation \eqref{eq:dyakov1} is a linear system in the three unknowns $s_1 ^{(\rm s \Plus)}$, $v_{1x}^{(\rm e \Plus)}$ and $z_1$. The term $\mathbb{b}$ represent the waves incident from upstream and it vanishes if the upstream fluid is unperturbed. In this latter case, the linear system has non-zero solutions only if
\begin{equation}
\det \mathbb{A} = 0. \label{eq:detA}
\end{equation}
Performing the calculations we obtain:
\begin{align}
& \det \mathbb{A} = \frac{i (\cs^2/v_0^2 - 1)}{2 k_y (k_x^{(\Plus)} v_0 - \omega)} \times \nonumber\\
& \left[ 
\left( 1 - \frac{v_0^2}{\cs^2} \right) (k_x^{(\Plus)} v_0 - \omega) ( \cs^2 k_y^2 + \omega^2) + 2 \omega ( k_y^2 v_0^2 + \omega^2)
\right], 
\end{align}
which coincides with the result of \cite{Dyakov54} and equation (90.10) of \cite{landau} except for an unimportant overall multiplication factor. We can now solve equation \eqref{eq:detA} coupled with equation \eqref{eq:dispdk1} (taken with the plus sign) in the two unknowns $\omega$, $k_x^{(\Plus)}$ to obtain the proper oscillation frequencies of the system:
\begin{equation}
\omega = \pm \cs k_y, \quad k_x^{(\Plus)} = \mp \frac{2 \cs k_y v_0}{\cs^2 - v_0^2}. \label{eq:resultDK}
\end{equation}
This is the result of \cite{Dyakov54} for our simple isothermal case. The eigenfrequencies of the system are real, meaning that the system is stable. This is also referred as ``spontaneous emission of waves'' from the shock \citep{landau}.

By solving the inhomogeneous case in which $\mathbb{b}\neq 0$ it is possible to obtain the transmission coefficients and the amplification factors of incident waves. The full formulas for these quantities can be found elsewhere and are not reported here \citep{Dyakov58a,Dyakov58b,Kontorovich58b,McKenzieWestphal1968}. Here we limit ourselves to mention that the expression for these quantities contain the quantity $\det \mathbb{A}$ in the denominator, and therefore they diverge if waves are sent with frequencies corresponding to the proper oscillation frequencies of the system. Thus, if sound waves spontaneously emitted from the shock were somehow allowed to re-enter from the other side, these could resonate with the shock, leading to unsteady flow. Large amplifications that are possible for these values of the frequency of incident waves provide a physical picture to explain why periodicity is the key that makes shocks unstable.

\section{An even simpler toy problem} \label{appendix:toy}

According to the interpretation given in the main text, shocks are always unstable under periodic boundary conditions, while they can be unstable under DK boundary conditions only if high shear in the post-shock region triggers a parasitic KHI, which we argued to be distinct from the periodic shock instability. To test whether this is true, we looked for a similar problem such that the steady state always has $v_{0y}=0$ (no shear). If our interpretation is correct, this problem should always be unstable under periodic boundary conditions, and always stable under DK boundary conditions. In this appendix we study such a variant of the main problem.

Consider the following:
\begin{align} 
	& \pa_t \bfv + \left( \bfv \cdot \nabla \right) \bfv  = - \frac{\nabla P}{\rho} -\nabla \Phi \label{eq:toyeom1}, \\ 
	& \pa_t \rho + \nabla \cdot \left( \rho \bfv \right) = 0 \, , \label{eq:toyeom2}
\end{align}
with 
\begin{equation}
\Phi(x) = \Phi_0 \cos\left( \frac{2 \pi x}{L} \right) - F x,
\end{equation}
where $F$ is a constant, and as before we assume the gas to be isothermal. This problem is equivalent to the problem posed by Eqs. \eqref{eq:eom1} and \eqref{eq:eom2} for the case $\Omega=F_y=0$. Dimensional analysis shows that without loss of generality we can put $F=L=1$, and the problem has only two dimensionless parameters: $\Phi_0$ and $\cs$.

\begin{figure}
\includegraphics[width=0.5\textwidth]{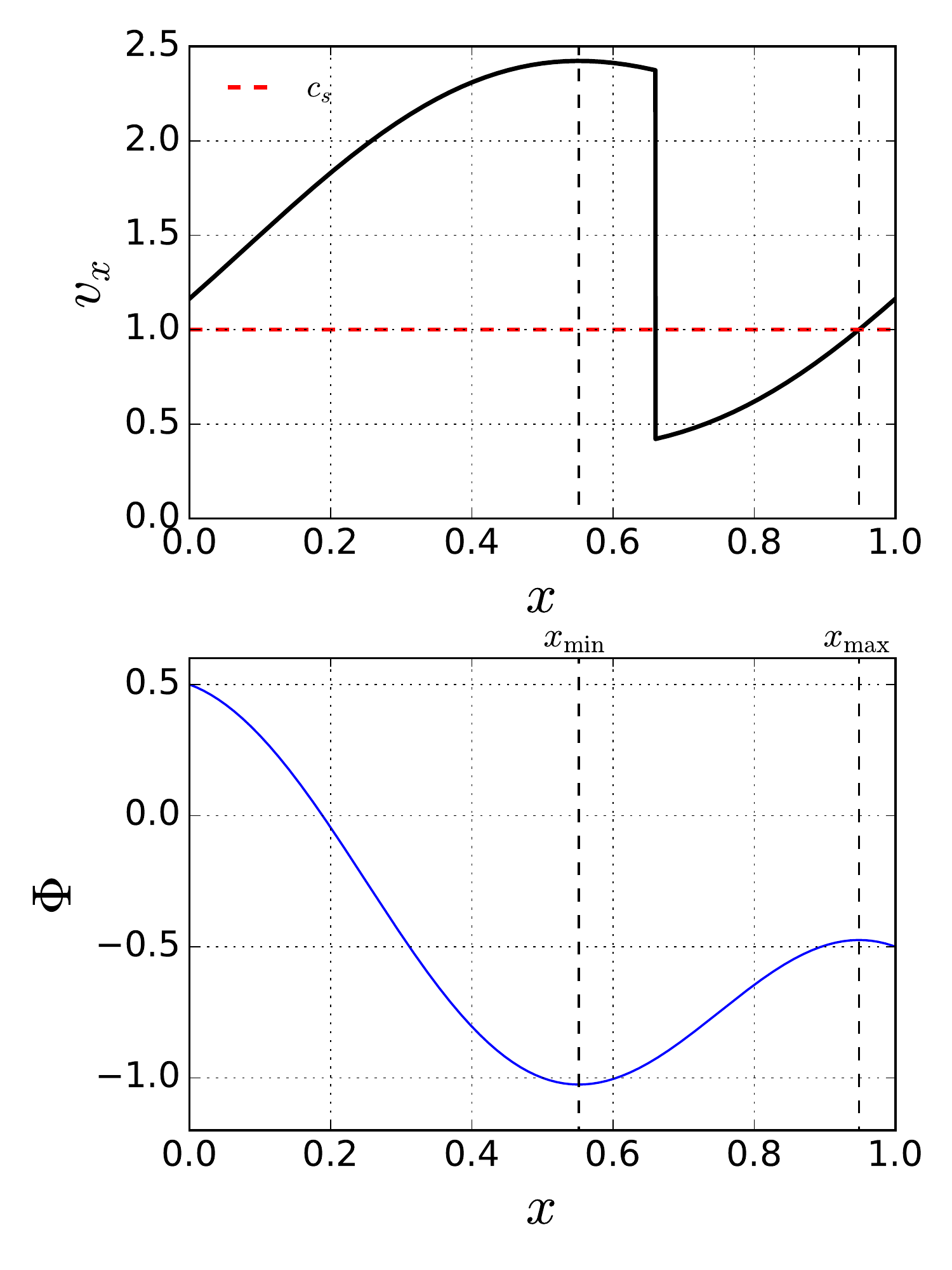}
 \caption{Top panel: an example of steady state solution for the toy problem for $\cs=1.0$, $\Phi_0=0.5$. Bottom panel: the corresponding $\Phi$.}
\label{fig:toy1} 
\end{figure}

Looking for steady states that depend only on $x$ and have period $L$ as before, we arrive at the following equations (which are the analog of Eqs. \ref{eq:s1a} and \ref{eq:s1b}) :
\begin{align}
v_{0x}' & = \frac{- \Phi'}{ v_{0x} - \frac{\cs^2}{v_{0x}}}, \label{eq:toys1b} \\
v_{0y}' & = 0. \label{eq:toys1b}
\end{align}
The top panel in Fig. \ref{fig:toy1} shows as an example the steady state solution obtained for $\cs=1$, $\Phi_0=0.5$, while the bottom panel shows the corresponding $\Phi$. In between shocks, fluid elements have a net gain of energy from the ever decreasing potential which is then radiated away at the shock, and the cycles starts over.

We have performed the same linear stability analysis that we presented in the main text. We used several values of $\cs=0.5,1.0$ and $\Phi_0=0.5,1.0,2.0$ in this toy problem. In every case, we have found that imposing periodic boundary condition the system is unstable, while imposing DK boundary conditions the system is stable.

Note that this problem is equivalent to the problem of nearly one-dimensional gas flow through a nozzle, where $\Phi(x)$ plays the role of the nozzle width $A(x)$ \citep[see for example \textsection 97 in][]{landau}. It is known that to accelerate gas from sub to supersonic velocities through a nozzle, the nozzle must be first converging and then diverging, and the sonic point occurs where the nozzle has minimum width. One cannot achieve supersonic velocities using an ever narrowing nozzle. Analogously, in our case the gas cannot achieve supersonic velocity through a monotonically decreasing $\Phi(x)$. $\Phi(x)$ must have a local maximum at the sonic point. This can also be seen from Eq. \eqref{eq:toys1b}: $v_{0x}' $ can remain finite at the sonic point only if $\Phi'=0$. Accelerating gas to supersonic velocities is a necessary to have shocks, so no solution with shocks can be found if maxima of $\Phi$ are not present. Note also that the equations and physical mechanism to accelerate gas from subsonic to supersonic velocities described here is essentially the same as in \cite{Parker1958,Parker1965} solar wind solution. In this latter case however the flow is not periodic but extends to infinity. 

\begin{figure}
\includegraphics[width=0.5\textwidth]{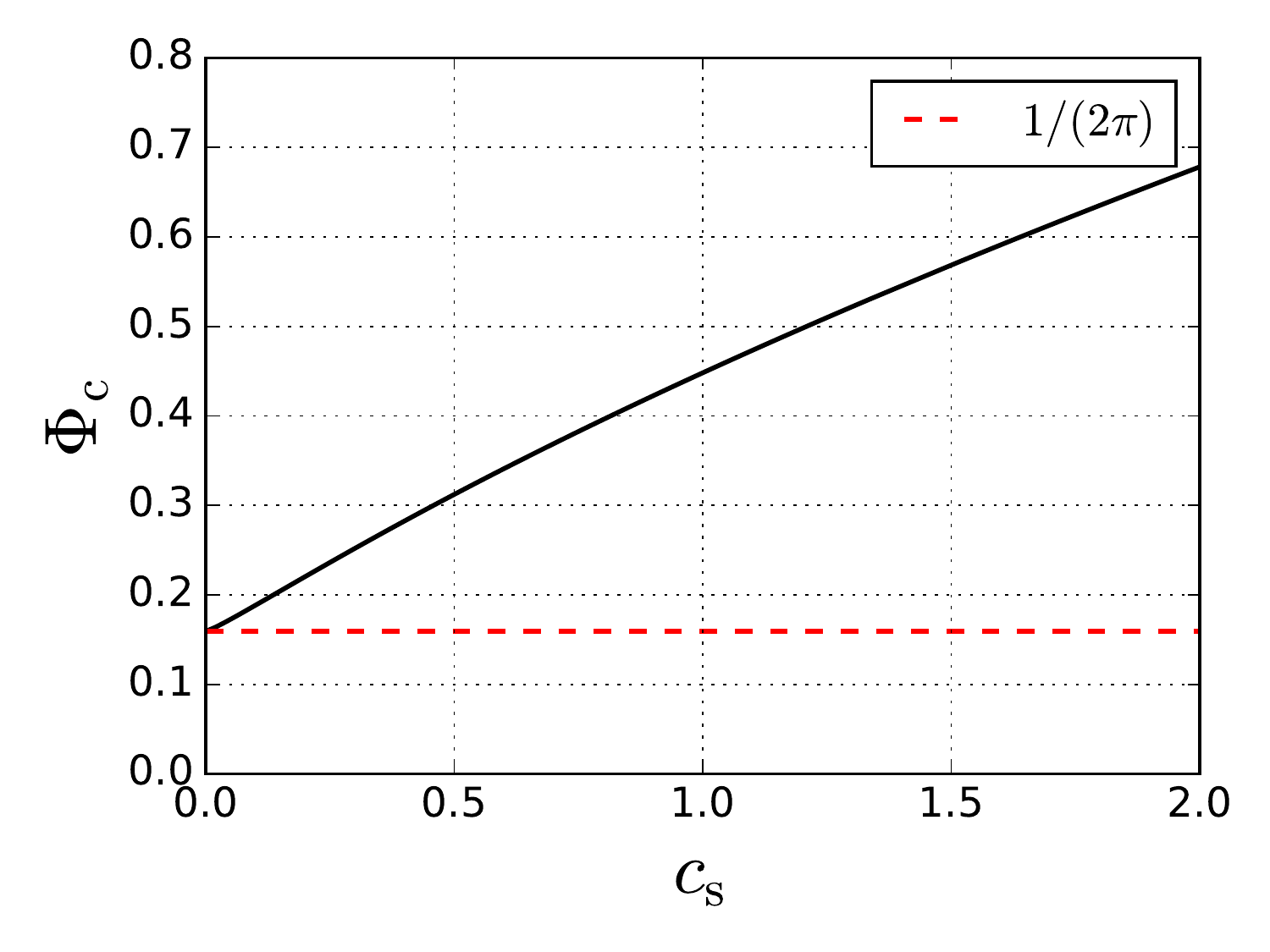}
 \caption{$\Phi_{\rm c}$ as a function of the only dimensionless parameter, $\cs$, for the toy problem.}
\label{fig:toyf} 
\end{figure}

Thus, the requirement that $\Phi(x)$ has local maxima puts a lower limit on $\Phi_0$:
\begin{equation}
\Phi_0 \geq \frac{F L } {2 \pi}. \label{eq:toylim1}
\end{equation}
It is tempting to identify this with $\Phi_{\rm c}$ (i.e., the minimum $\Phi$ for which shock solutions exist) for this toy problem. However this is a necessary but not sufficient condition to find a solution with periodic shocks in our problem. To understand why, consider Bernoulli's therorem, which states that between shocks the following quantity is conserved in our steady states:\footnote{At shocks in a periodic steady state it jumps by an amount $FL$.}
\begin{equation}
 \frac{v_x^2}{2} - \frac{\cs^2}{2} - \cs^2 \log\left(v_x/\cs\right) + \Phi(x) = \rm{constant}.
\end{equation}
Let $v_{\rm a}$ and $v_{\rm b}$ be the velocity just before and after the shock. If $\Phi_0$ is too small, for example just above the limit given by Eq. \eqref{eq:toylim1}, as we start integrating backwards from the sonic point (which coincides with $x_{\rm max}$), $v_x$ decreases until $x_{\rm min}$, but then it starts increasing again (look at how the sign of $v_x'$ depends on $\Phi'$ in Eq. \ref{eq:toys1b}). If $x_{\rm min}$ and $x_{\rm max}$ are too close, the flow starts decreasing before it has reached a sufficient velocity to satisfy the shock jump condition, that in this case is $v_{\rm a} v_{\rm b} = \cs^2$. A sufficient condition for the existence of solutions with periodic shocks can be found by imposing that when $\Phi_0=\Phi_{\rm c}$ the shock should appear at the position $x_{\rm min}$. This amounts to solving the following system in the three unknowns $v_{\rm a}$, $v_{\rm b}$, $\Phi_0$ for given $\cs$, $L$ and $F$:
\begin{align} 
	& \frac{v_{\rm a}^2}{2} - \frac{\cs^2}{2} - \cs^2 \log\left(v_{\rm a}/\cs\right) + \Phi(x_{\rm min}) = \Phi(x_{\rm max}) \\ 
	&  \frac{v_{\rm b}^2}{2} - \frac{\cs^2}{2} - \cs^2 \log\left(v_{\rm b}/\cs\right) + \Phi(x_{\rm min} + L) = \Phi(x_{\rm max}) \\ 
	& v_{\rm a} v_{\rm b} = \cs^2
\end{align}
The solution to this system yields a $\Phi_{\rm c}$ that depends on the only other dimensionless parameter of this toy problem:
\begin{equation}
\Phi_{\rm c} = \Phi_{\rm c}(\cs).
\end{equation}
This function can be calculated numerically by solving the system above and is shown in Fig. \ref{fig:toyf}. For $\cs=0.0$, the $\Phi_{\rm c}$ coincides with the lower limit \eqref{eq:toylim1}. This explains the origin of $\Phi_{\rm c}$ in this simple toy problem.

\section{Tables with solutions of the eigenvalue problem} \label{appendix:tables}

In this appendix we provide values of the initial conditions for $u_{1y}$ and eigenfrequencies for the modes with $k_y=0$ shown in Figs. \ref{fig:stability1} and \ref{fig:stability2}. The initial conditions are given at the sonic point, which is the point where we start integrating Eqs. \eqref{eq:main_ux}-\eqref{eq:main_rho} in our numerical scheme. The initial value of $s_1$ at the sonic point is always fixed to be $1+\imi$, while the initial value of $u_{1x}$ can be calculated from the sonic condition \eqref{eq:sonic}.

\clearpage

\begin {table}
\begin{center}
    \begin{tabular}{r r}
   \multicolumn{1}{c}{\Large $u_{1y}$} & \multicolumn{1}{c}{\Large $\omega$} \\[1mm]
    \hline
    $\Phi_0 =0.25$, $\cs=0.7$ &  periodic b.c. \\
    \hline
    $0.640 		+	0.640\imi$ 	&  	$0.000  	- 	1.670\imi$\\
    $0.279 		- 	0.837\imi$ 	&  	$2.567  	- 	0.361\imi$ \\
    $-4.331 	+ 	0.262\imi$ 	& 	$3.142  	+ 	0.000\imi$ \\
    $-1.995 	+ 	1.544\imi$ 	& 	$6.284  	- 	0.000\imi$ \\
    $0.169  	- 	0.267\imi$ 	& 	$8.624  	- 	0.001\imi$ \\
    $-4.009  	- 	2.907\imi$  	& 	$9.426 	+ 	0.000\imi$ \\
    $-3.613  	+ 	2.962\imi$ 	& 	$12.568  	- 	0.000\imi$ \\
    $-0.135  	+ 	3.665\imi$ 	& 	$15.710  	- 	0.000\imi$ \\
    $0.104  	- 	0.133\imi$ 	& 	$16.579  	- 	0.005\imi$ \\
    $-4.621  	+ 	3.751\imi$ 	& 	$18.852  	+ 	0.000\imi$ \\
    \hline
    $\Phi_0 =0.25$, $\cs=0.7$ &  DK b.c. \\
    \hline
    $0.838  	+ 	0.838\imi$ 	&  	$0.000 	 -	1.595\imi$ \\
    $0.433  	- 	0.678\imi$ 	&  	$3.110  	- 	1.137\imi$ \\
    $0.643  	- 	0.334\imi$ 	&  	$3.446  	- 	3.266\imi$ \\
    $0.481  	- 	0.310\imi$ 	&  	$4.608  	- 	3.085\imi$ \\
    $0.395 		- 	0.268\imi$ 	&  	$5.637  	- 	3.139\imi$ \\
    $0.329  	- 	0.232\imi$ 	&  	$6.772  	- 	3.218\imi$ \\
    $0.280  	- 	0.205\imi$ 	&  	$7.918  	- 	3.258\imi$ \\
    $0.244  	- 	0.184\imi$ 	&  	$9.066  	- 	3.286\imi$ \\
    $0.215  	- 	0.166\imi$ 	&  	$10.217  	- 	3.310\imi$ \\
    $0.192  	- 	0.152\imi$ 	&  	$11.371  	- 	3.330\imi$ \\
    $0.174  	- 	0.140\imi$ 	&  	$12.528  	- 	3.347\imi$ \\
    $0.158  	- 	0.130\imi$ 	&  	$13.686  	- 	3.363\imi$ \\
    $0.145  	- 	0.121\imi$ 	&  	$14.846  	- 	3.378\imi$ \\
    $0.134  	- 	0.113\imi$ 	&  	$16.007  	- 	3.392\imi$ \\
    $0.125  	- 	0.106\imi$ 	&  	$17.168  	- 	3.406\imi$ \\
    $0.117  		- 	0.100\imi$ 	&  	$18.329  	- 	3.419\imi$ \\
    $0.109  	- 	0.094\imi$ 	&  	$19.490  	- 	3.433\imi$ \\
    \hline
    \end{tabular}
\caption{Initial conditions at the sonic point and eigenfrequencies for modes with $k_y=0$ in Fig. \ref{fig:stability1}. \label{tab:1}}
\end{center}
\end{table}

\begin {table}
\begin{center}
    \begin{tabular}{r r}
   \multicolumn{1}{c}{\Large $u_{1y}$} & \multicolumn{1}{c}{\Large $\omega$} \\[1mm]
    \hline
    $\Phi_0 =0.025$, $\cs=0.3$ &  periodic b.c. \\
    \hline
    $0.000  	+ 	0.000\imi$ 	&  $0.000  	- 	0.000\imi$ \\
    $-0.770 	- 	0.860\imi$ 	&  $1.961  	+ 	0.005\imi$ \\
    $-0.322 	+ 	0.181\imi$ 	&  $3.143  	- 	0.000\imi$ \\
    $0.257  	- 	0.448\imi$ 	&  $6.085  	- 	0.003\imi$ \\
    $-0.057  	- 	0.196\imi$ 	&  $6.285  	- 	0.000\imi$ \\
    $-0.304  	+ 	5.976\imi$ 	&  $9.428  	+ 	0.000\imi$ \\
    $0.164  	- 	0.210\imi$ 	&  $10.990  	- 	0.001\imi$ \\
    $-4.037  	+ 	7.922\imi$ 	&  $12.571  	+ 	0.000\imi$ \\
    $-0.491  	+ 	0.798\imi$ 	&  $15.713  	- 	0.000\imi$ \\
    $0.116 		- 	0.136\imi$ 	&  $16.103  	- 	0.001\imi$ \\
    $-1.435  	+ 	2.412\imi$ 	&  $18.856  	- 	0.000\imi$ \\
    \hline
    $\Phi_0 =0.025$, $\cs=0.3$ &  DK b.c. \\
    \hline
    $-0.518 + 0.097\imi$	&  $0.776  - 0.868\imi$ \\
    $0.451  + 0.514\imi$ 	&  $0.823  - 3.033\imi$ \\
    $0.687  + 0.501\imi$ 	&  $1.844  - 3.172\imi$ \\
    $0.851  + 0.178\imi$ 	&  $2.977  - 3.204\imi$ \\
    $0.708  - 0.080\imi $	&  $4.109  - 3.251\imi$ \\
    $0.544  - 0.163\imi $	&  $5.231  - 3.294\imi$ \\
    $0.429  - 0.180\imi $	&  $6.348  - 3.329\imi$ \\
    $0.351  - 0.176\imi $	&  $7.462  - 3.359\imi$ \\
    $0.296  - 0.166\imi $	&  $8.576  - 3.386\imi$ \\
    $0.256  - 0.155\imi $	&  $9.689  - 3.411\imi$ \\
    $0.225  - 0.144\imi $	&  $10.801  - 3.435\imi$ \\
    $0.201  - 0.134\imi $	&  $11.911  - 3.459\imi$ \\
    $0.181  - 0.126\imi $	&  $13.019  - 3.483\imi$ \\
    $0.165  - 0.118\imi $	&  $14.123  - 3.506\imi$ \\
    $0.151  - 0.111\imi $	&  $15.223  - 3.526\imi$ \\
    $0.140  - 0.105\imi $	&  $16.321  - 3.541\imi$ \\
    $0.130  - 0.099\imi $	&  $17.418  - 3.550\imi$ \\
    $0.122  - 0.094\imi $	&  $18.520  - 3.555\imi$ \\
    $0.114  - 0.090\imi $	&  $19.626  - 3.558\imi$ \\
    \hline
    $\Phi_0 =0.25$, $\cs=0.3$ &  periodic b.c. \\
    \hline
    $- 0.000 - 0.000\imi$	& $0.000 + 0.000\imi$ \\
    $- 0.363  - 0.809\imi$ 	&  $1.759 + 0.237\imi$ \\
    $- 1.053  - 0.064\imi$	&  $3.143  + 0.000\imi$ \\
    $ 0.092  - 0.107\imi$ 	&  $6.285  - 0.000\imi$ \\
    $ 0.191  - 0.382\imi$ 	&  $6.674  + 0.023\imi$ \\
    $- 0.681  + 0.352\imi$ 	&  $9.428  - 0.000\imi$ \\
    $ 0.472  - 0.081\imi$ 	&  $12.571  + 0.000\imi$ \\
    $ 0.131  - 0.176\imi$ 	&  $12.967  - 0.108\imi$ \\
    $ 0.097  + 1.143\imi$ 	&  $5.713  - 0.000\imi$ \\
    $ 0.242  - 0.615\imi$	&  $18.856  + 0.000\imi$ \\
    $ 0.093  - 0.113\imi$	&  $19.371  - 0.103\imi$ \\
    \hline
    $\Phi_0=0.25$, $\cs=0.3$ &  DK b.c. \\
    \hline
   $0.574  + 0.574\imi$	&  $0.000  - 3.233 \imi $\\
    $-0.884  - 1.493\imi$	&  $1.701  - 1.067 \imi $\\
    $0.661  + 0.569\imi$ 	&  $2.621  - 4.864 \imi $\\
    $0.763  - 0.046\imi$ 	&  $3.375  - 4.046 \imi $\\
    $0.548  - 0.162\imi$ 	&  $4.840  - 4.110 \imi $\\
    $0.408  - 0.177\imi$ 	&  $6.319  - 4.158 \imi $\\
    $0.320  - 0.168\imi$ 	&  $7.808  - 4.191 \imi $\\
    $0.262  - 0.154\imi$ 	&  $9.305  - 4.216 \imi $\\
    $0.221  - 0.141\imi$ 	&  $10.805  - 4.236 \imi $\\
    $0.190  - 0.129\imi$ 	&  $12.309  - 4.252 \imi $\\
    $0.167  - 0.118\imi$ 	&  $13.813  - 4.267 \imi $\\
    $0.149  - 0.109\imi$ 	&  $15.320  - 4.280 \imi $\\
    $0.134  - 0.101\imi$ 	&  $16.828  - 4.293 \imi $\\
    $0.122  - 0.094\imi$ 	&  $18.338  - 4.306 \imi $\\
    $0.112  - 0.088\imi$ 	&  $19.848  - 4.321 \imi $\\
    \hline
    \end{tabular}
\caption{Initial conditions at the sonic point and eigenfrequencies for modes with $k_y=0$ in Fig. \ref{fig:stability2}. \label{tab:2}}
\end{center}
\end{table}

\end{document}